\documentstyle[emulateapj,psfig]{article}
\def\gs{\mathrel{\raise0.35ex\hbox{$\scriptstyle >$}\kern-0.6em \lower0.40ex\hbox{{$\scriptstyle \sim$}}}}
\def\ls{\mathrel{\raise0.35ex\hbox{$\scriptstyle <$}\kern-0.6em \lower0.40ex\hbox{{$\scriptstyle \sim$}}}}

\submitted{{\it Received:} November 30, 1998; {\it Revised:} January 18, 1999}

\begin{document}
\small

\title{The Star Formation Histories of Galaxies
in Distant Clusters\footnotemark}

\footnotetext{Based on observations 
obtained with the NASA/ESA Hubble Space Telescope
which is operated by STSCI for the Association of Universities
for Research in Astronomy, Inc., under NASA contract NAS5-26555.}

\author{
Bianca M.\ Poggianti,$^{\!}$\altaffilmark{2,3,4}
Ian Smail,$^{\!}$\altaffilmark{5}
Alan Dressler,$^{\!}$\altaffilmark{6}
Warrick J.\ Couch,$^{\!}$\altaffilmark{7}
Amy J.\ Barger,$^{\!}$\altaffilmark{8}\\
Harvey Butcher,$^{\!}$\altaffilmark{9}
Richard S.\ Ellis,$^{\!}$\altaffilmark{2}
\& Augustus Oemler, Jr.\altaffilmark{6}
}
\smallskip

\affil{\scriptsize 2) Institute of Astronomy, Madingley Rd, Cambridge CB3 OHA, UK}
\affil{\scriptsize 3) Royal Greenwich Observatory, Madingley Rd, Cambridge CB3 0EZ, UK}
\affil{\scriptsize 4) Osservatorio Astronomico di Padova, vicolo dell'Osservatorio 5, 35122 Padova, Italy}
\affil{\scriptsize 5) Department of Physics, University of Durham, South Rd, Durham DH1 3LE, UK}
\affil{\scriptsize 6) The Observatories of the Carnegie Institution of Washington, 813 Santa Barbara St., Pasadena, CA 91101-1292}
\affil{\scriptsize 7) School of Physics, University of New South Wales, Sydney 2052, Australia}
\affil{\scriptsize 8) Institute for Astronomy, University of Hawaii, 2680 Woodlawn Dr., Honolulu, HI  96822}
\affil{\scriptsize 9) NFRA, PO Box 2, 7990 AA Dwingeloo, The Netherlands}

\begin{abstract}
We present a detailed analysis of the spectroscopic catalog of galaxies
in 10 distant clusters from Dressler et al.\ (1999, D99).  We
investigate the nature of the different spectral classes defined by D99
including star forming, post--starburst and passive galaxy populations,
and reproduce their basic properties using our spectral synthesis
model.  We attempt to identify the evolutionary pathways between the
various spectral classes in order to search for the progenitors of the
numerous post--starburst galaxies.  The comparison of the spectra of
the distant galaxy populations with samples drawn from the local
Universe leads us to identify a significant population of
dust--enshrouded starburst galaxies, showing both strong Balmer
absorption and relatively modest [OII] emission, that we believe are
the most likely progenitors of the post--starburst population.  We
present the differences between the field and cluster galaxies at
$z=0.4$--0.5. We then compare the spectral and the morphological
properties of the distant cluster galaxies, exploring the connection
between the quenching of star formation inferred from the spectra and
the strong evolution of the S0 population discussed by Dressler et
al.\ (1997).  We conclude that either two different timescales and/or
two different physical processes are responsible for the spectral and
the morphological transformation.
\end{abstract}

\keywords{cosmology: observations --- galaxies: clusters: general ---
galaxies: evolution --- infrared: galaxies}

\section{Introduction}

Of all the environments explored locally, the dense, concentrated cores
of rich clusters are highly conspicuous for their lack of star-forming
galaxies. Thus it came as a surprise when Butcher \& Oemler (1978; see
also Butcher \& Oemler 1984) discovered a population of blue, possibly
star-forming, galaxies in the cores of rich clusters at $z\gs 0.2$.
Subsequent low and intermediate resolution spectroscopy of these
galaxies has confirmed they are cluster members and also that star
formation  is the cause of their blue colors (Dressler \& Gunn 1982,
1983, 1992; Lavery \& Henry 1986, 1988; Couch \& Sharples 1987, CS87; Henry
\& Lavery 1987; Soucail et al.\ 1988; Fabricant et al.\ 1991, 1994;
Couch et al.\ 1994, 1998; Abraham et al.\ 1996; Fisher et al.\ 1998).

While the variation in the blue population within rich clusters shows
evidence for dramatic evolutionary effects, additional signs  of change
were also uncovered by the spectroscopic studies.  These included a
class of galaxies, first identified by Dressler \& Gunn (1983), with no
detectable emission lines (and hence little or no on-going star
formation) but very strong Balmer absorption. Dressler \& Gunn inferred
the presence of a substantial population of A stars and concluded that
an epoch of strong star formation had abruptly ended in the recent
past, leaving the galaxy in a so--called `post-starburst' phase.  In
1987 Couch \& Sharples made the first attempt to synthesise a coherent
view of the star formation histories of these and the other spectral
populations within the clusters, hoping to identify a single
evolutionary cycle through a starburst phase to the post-starburst
galaxies and from there into the large population of passive galaxies
seen in the clusters (see also Barger et al.\ 1996).   However, the
nature of the physical process(es) responsible for triggering and
terminating the episodes of star formation in the distant clusters (and
the lack of such activity today) has remained elusive.
 
The next major step came from high resolution imaging with the {\it
Hubble Space Telescope} ({\it HST}\,).  Using pre- and
post-refurbishment imaging,  the populations of star-forming and
post-starburst galaxies were morphological identified with
disk-dominated systems, a fraction of which are interacting or
obviously disturbed (Couch et al.\ 1994, 1998;  Dressler et al.\ 1994;
Oemler et al.\ 1997).  This supported the suggestion of galaxy-galaxy
interactions as one triggering mechanism for the star formation seen in
the distant clusters (Lavery \& Henry 1988). However, the expectation
that the high-speed encounters within clusters would be ineffective at
disturbing galaxies has led to other dynamical mechanisms being
suggested, including tidal disruption (Byrd \& Valtonen 1990) and
repeated high--speed encounters in the cluster potential (`harassment',
Moore et al.\ 1996, 1998).  Interactions with the intracluster medium
(ICM)  have also remained a popular explanation of some of the
properties of the Butcher-Oemler population (Gunn \& Gott 1972).

The {\it HST} imaging of distant clusters also turned up evidence for
evolution in the passive cluster galaxies.  The population of luminous
cluster ellipticals appear to have been in place in the clusters since
at least $z\sim 0.6$ (Smail et al.\ 1997) and the homogeneity of their
properties would argue that they underwent the bulk of their star
formation prior to $z\sim 3$ (Bower, Lucey \& Ellis 1992; van Dokkum \&
Franx 1996; Ellis et al.\ 1997; Barger et al.\ 1998).  In contrast, the
S0 galaxies which dominate the cores of rich clusters today are
noticeably absent from these environments at $z\sim 0.5$ (Dressler et
al.\ 1997).  The recent formation or transformation of the cluster S0
galaxies and the connections to the evolutionary history of the
Butcher-Oemler population are of considerable interest for
understanding the extent of environmental influences on galaxy
morphology and stellar populations.

To elucidate the physical processes which are driving the star
formation activity and the morphological transformations in these
distant clusters, we have been conducting a  large space- and
ground-based study of 10 rich clusters in the range  $0.37\leq z\leq
0.56$. One of the main aims of this study -- known as the `MORPHS'
collaboration -- has been to obtain detailed morphologies of
magnitude-limited  samples of galaxies in these clusters from high
resolution images taken  with the Wide Field and Planetary Camera 2
(WFPC2) onboard the {\it HST}. These data have already been presented
and discussed in a series of papers (Smail et al.\ 1997a,b; Ellis et
al.\ 1997; Dressler et al.\ 1997; Barger et al.\ 1998).

Another important component of our program has been to obtain extensive
ground-based spectroscopy within the fields of these clusters in order
to better quantify the star formation activity of the cluster galaxies,
particularly those for which {\it HST} morphologies are available. In
Dressler et al.\ (1999; hereafter D99), we presented spectra for 657
galaxies observed across the fields of our 10 clusters, 424 of which
were   confirmed to be cluster members.  Detailed morphological and
photometric information  from the {\it HST} images (Smail et al.\ 1997b,
S97) is available for 204 cluster members and 71 field galaxies and has
been included in the D99 tables.  This dataset is suitable to study
simultaneously the spectral and  the morphological properties of
galaxies in distant clusters and it allows a direct comparison between
the high redshift field  and cluster populations.  The main results of
D99 can be summarized as follows:

1) The `post-starburst' (k+a/a+k) galaxies are a significant fraction,
$\sim 20$\%, of the cluster population, while their incidence in the
field at similar redshifts is considerably lower.  Moreover, the
frequency of k+a/a+k's in clusters is much higher at $z=0.4$--0.5 than
at low redshift, and apparently evolves more strongly with redshift
than in the field population.  While the galaxies in this class include
examples of all Hubble types, the majority of them show disk dominated
morphologies.  

2) At a given Hubble type, the cluster galaxies show a higher frequency
of spectra with low or no detectable [O{\sc ii}] emission compared to
the surrounding field (including examples of late-type spirals with no
current star formation).  This indicates a general reduction in the
star formation activity of galaxies in the clusters, as measured by the
EW([O{\sc ii}]), compared to the field.

3) Those galaxies which are currently forming stars (all the spectral
classes with emission lines) have both a more extended spatial
distribution and a higher velocity dispersion than the galaxies with
passive (k--type) spectra.  The recently star forming k+a/a+k galaxies
exhibit spatial and kinematic distributions intermediate between the
passive and active populations.  Dressler et al.\ (1999) interpret the
kinematic and evolutionary behaviour of the post-starburst galaxies as
clear evidence for the environment being responsible for the formation
or visibility of the k+a/a+k classes.

D99 gave a qualitative discussion of the properties of the cluster
galaxies and laid the ground-work for the quantitative analysis
undertaken in this work.  The purpose of this paper is to model and
interpret the different spectroscopic classes in D99 in order to
identify the star formation patterns that are present in cluster
galaxies at these  earlier epochs and thus predict their likely
subsequent (and also previous) spectrophotometric evolution.  This
information is vital to detailing and therefore understanding the
physical processes responsible for the changes in luminosity,  color
and morphology that underlie the `Butcher--Oemler' effect. It is worth
stressing that, although the existence of a strong evolution between
$z\sim 0.4-0.5$ and $z=0$ is well established, a dataset equivalent to
the one presented in D99 does not exist for low redshift clusters and
this of course limits the comparison with the present epoch.

The plan of this paper is as follows: in \S2 we describe the different
spectroscopic classes found in the distant clusters and examine the
possible selection effects that influenced their inclusion in the
spectroscopic sample. In \S3, we model and interpret each spectroscopic
class separately; in this section we also examine whether analogous
systems exist in the local Universe and what these can tell us about
the effects of dust extinction on the spectral properties of starburst
galaxies.  In \S4 we analyse the predicted range of spectrophotometric
paths of each class and investigate the possible evolutionary
connections among the classes.  We then use our sample of spectra for
similarly distant `field' galaxies to compare the star formation
activity in and out of clusters at this epoch, and we combine what has
been learned on the star formation activity of the different classes
with their {\it HST} morphologies, looking for connections between the
two.   In \S5 we also examine whether star formation activity in
distant clusters is correlated  with global cluster properties and we
then summarize our main conclusions in \S6.  Unless otherwise stated we
assume $q_0 = 0.5$ and take $h = H_0/100$\,km\,s$^{-1}$\,Mpc$^{-1}$.

\section{The spectroscopic catalog}

In this section we briefly outline the spectral classification scheme
adopted in D99 and we analyse the possible selection effects in this
sample.  Our classification scheme is based primarily on two lines,
[O{\sc ii}]$\lambda$3727 and  $\rm H\delta$($\lambda = 4101$ \AA),
which are good indicators of (respectively) current and recent star
formation (SF) in distant galaxy spectra. All the equivalent widths in
this paper are given in \AA\ in the rest frame and with a negative
value when in emission.
  
The spectral classification scheme is presented in Table~1, taken from
D99.  A full explanation of the notation denoting the various spectral
classes can be found in D99.  Only 62 out of the 657 spectra were too
poor to be assigned a spectral class, and in 38 of these at least one
emission line could be identified.  Thus we have spectral
classifications for around 95\% of our sample. 

For the purposes of this paper we have also divided the Balmer-strong
galaxies with no emission (k+a/a+k type) into `blue' and `red' galaxies
(as was done in CS87 for the PSG and HDS classes respectively); in this
case the color threshold has been chosen to be $(g-r)=1.15$ (in the
observed frame at $z=0.4$), equivalent to the one adopted in CS87 ($B_J
- R_F =2$ at $z=0.31$), and it corresponds to the separation between
the expected color of the early-type galaxies (Sa or earlier) from the
later-type spirals and irregulars (Sb and  later).  The Galactic
extinction in the direction of the clusters in our sample is negligible
(S97), except in the case of Cl\,0303$+$17 ($E(B-V)=0.12$),  and no
reddening correction has been applied in this paper.

\begin{table*}
{\scriptsize
\begin{center}
\centerline{\sc Table 1}
\vspace{0.1cm}
\centerline{\sc Spectral Classification Scheme from D99}
\vspace{0.3cm}
\begin{tabular}{lcccl}
\hline\hline
\noalign{\smallskip}
 {Class} & {EW [O{\sc ii}]\,3727} & {EW H$\delta$} &  Color & Comments \cr
 & (\AA) & (\AA) & &  \cr
\hline
\noalign{\smallskip}
k & absent & $< 3$ & ... & passive \cr
k+a & absent & 3--8 & ... & moderate Balmer absorption without emission \cr
a+k & absent & $\geq 8$ & ... & strong Balmer absorption without emission \cr
e(c) & yes,$< 40$ & $< 4$ & ... & moderate Balmer absorption plus emission, spiral-like \cr
e(a) & yes & $\geq 4$ & ... & strong Balmer absorption plus emission \cr
e(b) & $\geq 40$ & ... & ... & starburst \cr
e(n) & ... & ... & ... & AGN from broad lines or [O{\sc iii}]\,5007/H$\beta$ ratio \cr
e &  yes & ? & ... & at least one emission line, but S/N too low to classify \cr
? &   ?  & ? & ... & unclassifiable \cr
\noalign{\smallskip}
CSB & ... & ... & very blue & photometrically-defined starburst \cr 
\noalign{\smallskip}
\noalign{\hrule}
\noalign{\smallskip}
\end{tabular}
\end{center}
}
\vspace*{-0.8cm}
\end{table*}

\setcounter{table}{1}

The Balmer strong population is evident in Fig.~1 which shows the
EW($\rm H\delta$) as a function of the observed $(g-r)$ color for
cluster and field galaxies.  Such a diagram was presented for the first
time for a large spectroscopic  dataset in three clusters at $z=0.31$
by CS87.  The k+a/a+k galaxies stand out as a numerous population both
at red and blue colors (filled dots with EW($\rm H\delta)> 3$\AA), as
seen in previous  spectroscopic surveys. e(a) galaxies (crosses with
EW($\rm H\delta)> 4$\AA) dominate the blue side of the diagram,
together with the e(c) galaxies (most of the crosses with EW($\rm
H\delta)< 4$\AA).  The k class, represented by the filled dots with
EW($\rm H\delta)< 3$ \AA, have red colors consistent with their passive
spectrum.  The difference in Fig.~1 between the cluster and the field
diagrams is striking, and we will discuss it in \S5.1; the field is
missing the large population of k+a/a+k galaxies, while a large number
of e(a) spectra are still present (D99).

\hbox{~}
\vspace{-1.6in}
\centerline{\hspace*{2cm}\psfig{file=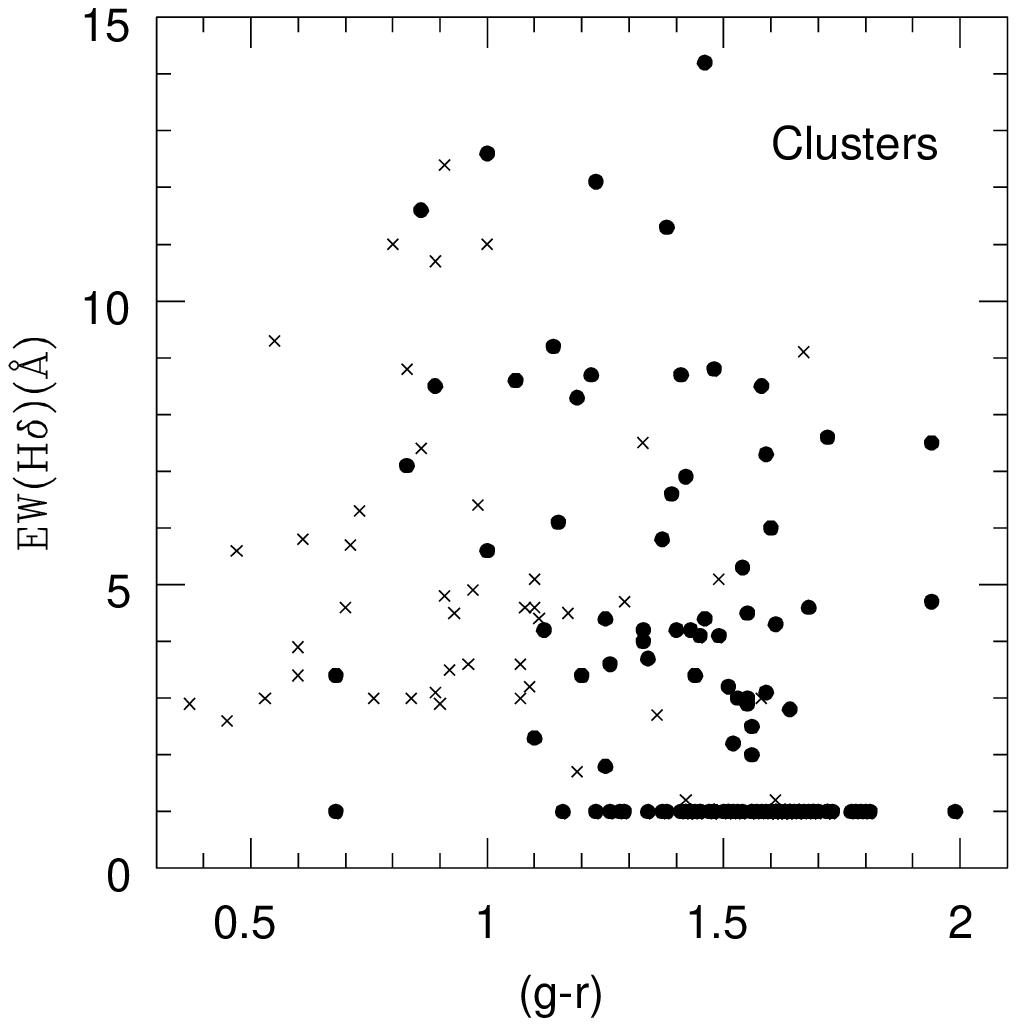,angle=0,width=4.0in}}
\vspace*{-1.8in}
\centerline{\hspace*{2cm}\psfig{file=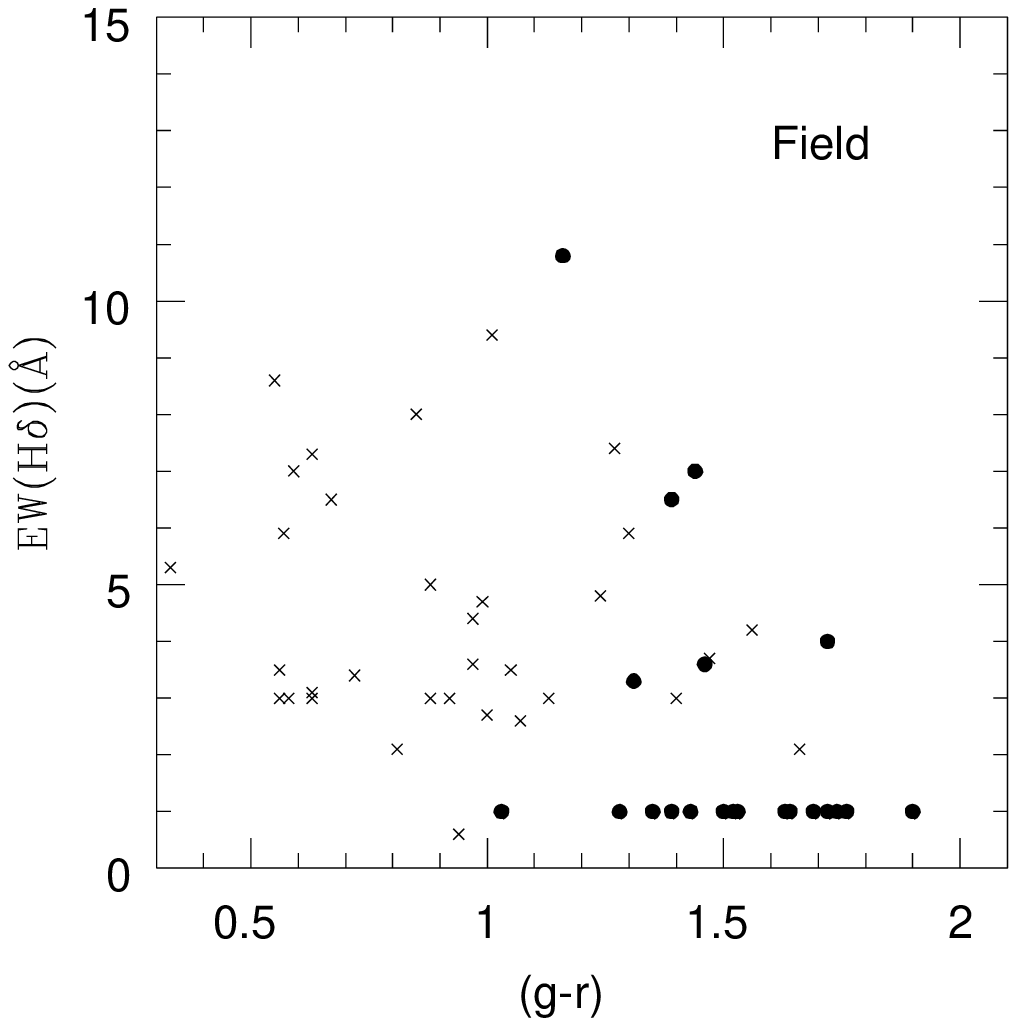,angle=0,width=4.0in}}
\vspace{-0.2in}

\noindent{\scriptsize
\addtolength{\baselineskip}{-3pt} 
\hspace*{0.1cm} Fig.~1.\ Color--H$\delta$ diagrams for the cluster
(top) and field (bottom) populations from D99.  Crosses and circles
indicate spectra with and without emission lines, respectively.  Only
members with $r<22.5$ are presented in the plot.  `e(n)', `e' and
uncertain (:) spectral classes are excluded.  If not measured, the
EW($\rm H\delta$) of k type galaxies have been  set equal to 1\AA.  The
$(g-r)$ color has been {\sc K}--corrected to  $z=0.4$. 

\addtolength{\baselineskip}{3pt}
}

\subsection{Sample selection}

The selection of the spectroscopic targets was influenced by galaxy
morphology within the core region covered by the WFPC2, while outside
this area the catalog is essentially a magnitude-limited sample to $i
\sim 21$ mag (D99).  The comparison between D99 and the morphological
catalog (S97) allows us to quantify this effect and apply the
appropriate corrections. For each cluster we compared the fraction of a
given morphological type in the spectroscopic catalog (including both
cluster members and field galaxies in each cluster field) to the
proportion observed in the {\it HST} catalog (S97) down to the limiting
magnitude in D99.
 
\hbox{~}\smallskip
{\scriptsize
\begin{center}
\centerline{\sc Table 2}
\vspace{0.1cm}
\centerline{\sc Morphological correction factors}
\vspace{0.3cm}
\begin{tabular}{lccc}
\hline\hline
\noalign{\smallskip}
 {Cluster} & E & S0 & Sd/Irr \cr
\hline
\noalign{\medskip}
Cl\,1447$+$23 & 1.43 & --   &  --  \cr
Cl\,0024$+$16 & 1.76 & 1.83 & 0.42 \cr
Cl\,0939$+47$ & 1.26 & 1.62 & 0.93 \cr
Cl\,0303$+17$ & 1.31 & 1.09 & 0.39 \cr
3C\,295       & 2.79 & 1.74 & 0.51 \cr
Cl\,1601$+$42 & 1.19 & 2.18 &  --  \cr
Cl\,0016$+$16 & 1.88 & 2.55 & 0.46 \cr
\noalign{\smallskip}
Average clusters       & 1.61 & 1.75 & 0.60 \cr
\noalign{\smallskip}
Field         & 1.55 & 1.91 & 0.34 \cr
\noalign{\smallskip}
\noalign{\hrule}
\noalign{\smallskip}
\end{tabular}
\end{center}
}

The ratio of these two fractions gives the total (cluster + field)
morphological correction factors. These  have been subsequently
partitioned between cluster and field galaxies to find the final
morphological corrections we have applied to the distributions in D99.
The number of galaxies of a given Hubble type in the spectroscopic
catalog needs to be multiplied by these factors, given in Table~2, in
order to find the ``true'' number which would have been observed in an
unbiased sample.  The table shows that, for the inner regions of the
cluster, the Sd/Irr galaxies are slightly oversampled relative to the
Sa--Sc types, while the E/S0 population is slightly undersampled. The
values in Table~2 are normalized taking the correction factor for Sa,
Sb and Sc galaxies equal to 1; the morphological types have been
classed according to the following T types:  E(T types from $-6$ to
$-4$ both included), S0($-3$,0), Sa(1,2), Sb(3,4), Sc(5,6),
Sd/Irr(7,10).  Three fields have not been included in this table:
Cl\,0054$-$27 and Cl\,0412$-$65, for which the number of good quality
spectra is too small, and the outer field in A\,370.  To estimate the
correction to the spectral classes we then weight these factors by the
observed distribution of spectral classes within each morphological
type (Table~3). 

\begin{table}
{\scriptsize
\begin{center}
\centerline{\sc Table 3}
\vspace{0.1cm}
\centerline{\sc Fraction of Galaxies of a given Hubble type
as a Function of Spectral Class}
\vspace{0.3cm}
\begin{tabular}{lcccccccccc}
\hline\hline
\noalign{\smallskip}
 {Morphology} & N$_{\rm tot}$ & k & k+a & a+k & e(a) & e(c) & e(b) & e(n) & e & ? \cr
\hline
\noalign{\medskip}
\multispan{11}{\hfil{Cluster~Sample}\hfil}\cr
\noalign{\smallskip}
 E &   54 & 0.81$\pm$0.12 & 0.09$\pm$0.04 & 0.02$\pm$0.02 & 0.02$\pm$0.02 & 0.04$\pm$0.03 & 0 & 0.02$\pm$0.02 & 0 & 0 \cr 
 S0 &   24 & 0.62$\pm$0.16 & 0.21$\pm$0.09 & 0.04$\pm$0.04 & 0 & 0.08$\pm$0.06 & 0 & 0 & 0 & 0.04$\pm$0.04 \cr 
 Sa &   39 & 0.69$\pm$0.13 & 0.13$\pm$0.06 & 0.03$\pm$0.03 & 0.05$\pm$0.04 & 0.05$\pm$0.04 & 0 & 0 & 0 & 0.05$\pm$0.04 \cr 
 Sb &   38 & 0.29$\pm$0.09 & 0.21$\pm$0.07 & 0.08$\pm$0.05 & 0.13$\pm$0.06 & 0.13$\pm$0.06 & 0 & 0.05$\pm$0.04 & 0.08$\pm$0.05 & 0.03$\pm$0.03 \cr 
 Sc &   29 & 0.17$\pm$0.08 & 0.17$\pm$0.08 & 0.10$\pm$0.06 & 0.28$\pm$0.10 & 0.10$\pm$0.06 & 0.10$\pm$0.06 & 0.03$\pm$0.03 & 0.03$\pm$0.03 & 0 \cr 
 Sd/Irr &   20 & 0.10$\pm$0.07 & 0.05$\pm$0.05 & 0.05$\pm$0.05 & 0.25$\pm$0.11 & 0.15$\pm$0.09 & 0.35$\pm$0.13 & 0 & 0 & 0.05$\pm$0.05 \cr 
\noalign{\medskip}
\multispan{11}{\hfil{Field~Sample}\hfil}\cr
\noalign{\smallskip}
 E &   ~9 & 0.44$\pm$0.22 & 0 & 0 & 0.22$\pm$0.16 & 0.22$\pm$0.16 & 0 & 0 & 0.11$\pm$0.11 & 0 \cr 
 S0 &   ~6 & 0.33$\pm$0.24 & 0.33$\pm$0.24 & 0 & 0 & 0 & 0.33$\pm$0.24 & 0 & 0 & 0 \cr 
 Sa &   ~6 & 0.67$\pm$0.33 & 0 & 0 & 0 & 0.33$\pm$0.24 & 0 & 0 & 0 & 0 \cr 
 Sb &   11 & 0.27$\pm$0.16 & 0 & 0 & 0.09$\pm$0.09 & 0.36$\pm$0.18 & 0.09$\pm$0.09 & 0 & 0.09$\pm$0.09 & 0.09$\pm$0.09 \cr 
 Sc &   26 & 0.04$\pm$0.04 & 0.04$\pm$0.04 & 0 & 0.08$\pm$0.05 & 0.54$\pm$0.14 & 0.19$\pm$0.09 & 0 & 0.08$\pm$0.05 & 0.04$\pm$0.04 \cr 
 Sd/Irr &   13 & 0 & 0 & 0 & 0.15$\pm$0.11 & 0.23$\pm$0.13 & 0.15$\pm$0.11 & 0 & 0.38$\pm$0.17 & 0.08$\pm$0.08 \cr 
\noalign{\smallskip}
\noalign{\hrule}
\noalign{\smallskip}
\end{tabular}
\end{center}
}
\vspace*{-0.8cm}
\end{table}

In the same way we have checked for any bias in  the spectroscopic
catalog related to the degree of distortion or disturbance of the
galaxy's structure, as measured by the disturbance index $D$ (S97).
Such a bias is not present in our sample and after having corrected for
the morphological bias in the central fields,  the whole spectral
dataset represents an essentially magnitude-limited sample.  All the
results presented in this work have been corrected for these selection
effects.

We note that the spectroscopic sample analysed here is roughly limited
at an absolute magnitude of $M_V=-19+5 \log_{10} h$ or $M_V^\ast+1.5$
(D99) while the morphological catalog on which our sample is based
reaches down to $M_V^\ast+3.5$; this itself is 3 magnitudes brighter
than the 5\,$\sigma$ detection limit of the {\it WFPC2} imaging data at
the median cluster redshift (S97).  The continuity in the sample
properties for the galaxies in the full (deeper) morphological catalog
compared to the spectroscopic sub-sample suggests that the latter is
not missing large classes of galaxies due to surface-brightness
selection limits, although we cannot rule out a subtle bias resulting
from the original object selection.

\section{The interpretation of the spectral classes}

In this section we model and interpret each spectroscopic class in
isolation, before attempting to establish possible evolutionary
relations  between the various classes (\S4).  In some cases we will
also examine the properties of low redshift galaxies with similar
spectra to provide further insights into the physical processes which
are signposted by particular combinations of spectral features.  In all
of this our aim is to constrain, on the basis of the observed spectra,
the past, present (and future) star formation history of these
galaxies. 

All the models employed in this paper have been generated with the
spectrophotometric code of Barbaro \& Poggianti (1997).  This code has
a number of advantages for the purpose of our study, in particular it
includes both the contribution of the stellar component and the thermal
emission of  the ionized gas in H{\sc ii} regions.

The stellar component is computed by an evolutionary synthesis model
which takes into account the main evolutionary phases up to the AGB
and post-AGB.  The stellar model allows us to compute the integrated
Spectral  Energy Distribution (SED) at low resolution over a wide
wavelength range (20\AA, 1000--25000\AA) and at higher resolution across
a shorter interval (4\AA, 3500--7500\AA). The first version, based
on Kurucz stellar atmosphere models (Kurucz 1993), is suitable for
computing colors  and studying low-resolution features (such as the
4000\AA\ break) with a wide wavelength coverage and including
stellar populations of different metallicities.  The second version,
based on the stellar spectral library observed by Jacoby et al.\ (1984)
(mostly of solar metallicity),  has enough resolution to study
absorption features, such as the stellar lines of the Balmer series.

In studying star-forming galaxies it is particularly useful  to
reproduce the nebular emission lines, which are the most prominent
features of the integrated spectrum and which are directly related to
the current star formation rate.  The Kurucz--based version of the
stellar model computes the ionizing flux  (below 912\AA) which  is used
to derive the luminosities of the nebular Balmer lines assuming case  B
recombination (optically thick,  Osterbrock 1989). The luminosities of
a number of metallic lines,  including [O{\sc ii}]$\lambda$3727, are
computed using the grid of H{\sc ii} region models given by Stasinska
(1990). The emission line spectrum is obtained  assuming solar
metallicity gas, and all the model equivalent widths have been measured
using the same software that was used on the data.

Figure 2 shows the various types of star formation histories we have
considered, which include spiral--like models (panel a), truncated
models (panel b), truncated models with a burst (panel c) and
spiral--like models with a burst (panel d).  The SFRs and properties of
the spiral--like models have been published in Barbaro \& Poggianti
(1997) and resemble the set of SF histories described by Sandage
(1986), where the ratio of present to past average SF increases towards
later types. They correspond to the Bruzual \& Charlot $\mu$ models
with a wide range of SF timescales ($\tau$); in particular, we have
investigated a set of models with $\tau$ between 0.5 Gyr and $\infty$
(constant SFR), including also a model whose SFR constantly increases
with time.

\subsection{The k+a/a+k class -- a review}

A large amount of theoretical work has been devoted to the
interpretation of  k+a/a+k spectra (Dressler \& Gunn  1983; CS87; Henry
\& Lavery 1987; Newberry et al.\ 1990; Belloni et al.\ 1995; Barger et
al.\ 1996;  Poggianti \& Barbaro 1996; Morris et al.\ 1998).  The main
conclusions can be summarized as follows:

a) A spectrum with strong Balmer lines in absorption (EW$(\rm H\delta)>
3$\AA) and negligible emission lines belongs to a galaxy which has no
significant current SF, but which was forming stars in the recent past
($< 1.5$ Gyr). Strong $(\rm H\delta)$ equivalent widths (${\rm
EW(H}\delta) >$4--5\AA) can only be reproduced  by models seen in a
quiescent phase soon after a starburst; for this reason k+a/a+k spectra
are often identified with  `post-starburst galaxies'.  Moderately
strong Balmer lines (${\rm EW(H}\delta)<$4--5\AA) can also be obtained
by simply halting the star formation in a galaxy which was forming
stars in a regular, continuous manner (truncated models), but the exact
threshold between these two regimes is not well defined, since it is
affected by the uncertainties in the measurement of the EW, both in the
data and in the models.  Newberry et al.\ (1990) showed that truncated,
as well as burst models, could reproduce the properties of passive,
Balmer-strong galaxies seen in the first spectroscopic surveys, which
gave much looser constraints than subsequent data.  More recently,
truncated models without a starburst are advocated by Morris et
al.\ (1998), although their definition of a 'truncated model' is
different from the one generally adopted.  In fact, they consider a 1
Gyr period of constant star formation, followed by a complete cessation
of the SF; such a model reproduces high EW$(\rm H \delta$) during a
period of 1--2 Gyr after the halting of the SF.  In this model the
galaxy did not form any star before the 1 Gyr constant SF (1--3 Gyr
before the $\rm H \delta$ strong phase) and using the definitions
commonly employed in the literature, this is a 'post--starburst' model
where the burst involves 100\% of the galactic mass. 

\hbox{~}
\vspace{-0.2in}
\centerline{\hspace{1.7in}\psfig{file=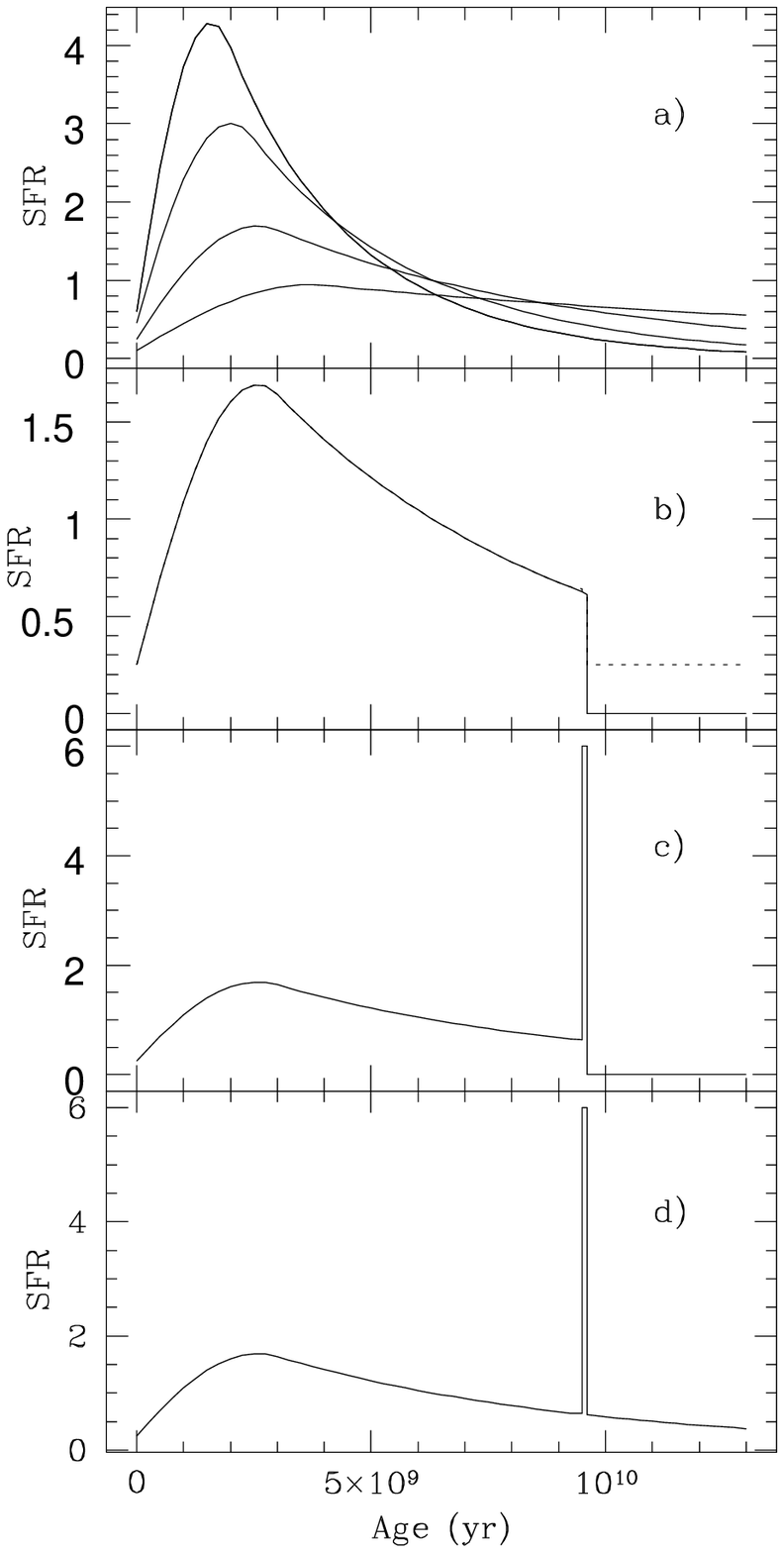,angle=0,width=3.5in}}
\noindent{\scriptsize
\addtolength{\baselineskip}{-3pt} 
\hspace*{0.3cm} Fig.~2.\ Star formation histories of the various
classes of models. SF rates are in arbitrary units: a) spiral-like
models (Barbaro \& Poggianti 1997), simulating an `Sa' (upper curve at
zero age),  `Sb', `Sc' and `Sd' galaxy (lower curve).  The model
parameters were determined by requiring the SED to reproduce at an age
of 16\,Gyr the average observed colors and gas fraction of present day
galaxies (Barbaro \& Poggianti 1997); the chemical evolution was
computed with a simple model with inflow that assumes the SFR to be
proportional to the gas fraction.  b) truncated models. The solid line
shows the spiral-like `Sc' model where the star formation is suddenly
halted.  The dotted line represents the case in which some residual SF
remains after the abrupt decrease. c) truncated models with burst.  The
spiral-like `Sc' model, after a strong burst lasting 0.1\,Gyr,  has no
further SF. d) a burst is superimposed on a spiral-like (`Sc') SF
history. At the end of the burst, a regular SF is resumed.

\addtolength{\baselineskip}{3pt}
}
\smallskip

b) The last star-formation event ended between a few Myr and 1.5\,Gyr
before the time of observation.  This range of timescales is determined
by the lifetime of the stars responsible for the strong Balmer lines
(Poggianti \& Barbaro 1997);

c) Spectral models using a standard Salpeter IMF show that a
significant fraction of the galactic mass needs to be involved ($>
10$--20\%) in the recent star formation episode in order to explain the
highest EW($\rm H\delta$) typically observed. In fact, the most extreme
cases  (${\rm EW(H}\delta)>10$\AA) can hardly be reproduced with a
standard IMF at all (Poggianti \& Barbaro 1996; Zabludoff,
priv.\ comm.).
   
d) The duration of any previous starburst phase is usually considered
unconstrained, with the exception of the work of Barger et
al.\ (1996) which found possible support for short-lived bursts ($\sim
0.1$\,Gyr);

e) The SF history {\it prior} to any burst cannot be determined from a
post-starburst spectrum (unless ultraviolet data sampling around
1500\AA\ in the rest frame is available). However, the high mass
fractions implied for the burst have always suggested that later-type
galaxies are more likely progenitors owing to their higher gas
fractions (Poggianti 1994). This suggestion has received support from
the  {\it HST}-based morphological studies (Couch et al.\ 1994, 1998;
Dressler et al.\ 1994, 1998; Oemler et al.\ 1997; S97), which typically
show that the morphologies of the post-starburst population are
predominantly disk-dominated.

f) The spectral properties {\it after} the k+a/a+k phase depend of
course on the subsequent SF history: if no SF is resumed, the signature
of the latest star formation episode (i.e.\ the strong $\rm H\delta$)
eventually disappears around 1--1.5\,Gyr after the end of the star
formation.  After this the galaxy shows a k type spectrum.

g) All the spectroscopic surveys, including D99, identify a group of
galaxies with very strong $\rm H\delta$ {\it and} very red colors.
These remain unexplained by the spectral models, which predict that the
galaxies with strong $\rm H\delta$ should remain moderately blue, even
as they decline to weaker $\rm H\delta$ strengths. For example, in
Fig.~1 they are the 13 cluster  galaxies with $(g-r)>1.3$ and $\rm
EW(H\delta)>5$\AA, lying in a part of the H$\delta$--$(g-r)$ plane
which cannot be reached by any simple post--starburst model (CS87;
Poggianti \& Barbaro 1996; Barger et al.\ 1996; Morris et al.\ 1998).
This discrepancy could be due in part to uncertainties in the
zero--point matching of the observational and theoretical photometric
systems (presumably 0.1--0.2 mag at most), as well as the result of
observational errors.  Nevertheless, it is unlikely these effects are
responsible for all the galaxies observed in this region of the
diagram.  Couch \& Sharples (1987) and Morris et al.\ (1998) discuss
possible solutions to the problem, which include internal reddening,
super-solar metallicities, or a top-heavy IMF for the stars produced
during the recent star formation.  However, a definitive conclusion has
not been reached so far, and we will address this issue later in \S4.

\subsection{The e(b) class}

Our models show that EW([O{\sc ii}])=40\AA\ cannot be reached with a
spiral--like star formation history. We have explored the set of star
formation histories shown in Fig.~2a and even models with constant or
constantly rising SFR do not reach EWs$>40$\AA, which seem to require a
burst of star formation.  However it should be kept in mind that the
emission models are for solar metallicity H{\sc ii} regions,  and that
the relation between SFR and [O{\sc ii}] flux changes substantially
with metallicity.  Although their strong emission lines testify to a
high current SFR, the interpretation of these galaxies as `starbursts'
is not straightforward:  the [O{\sc ii}] equivalent widths of the e(b)
galaxies have values comparable to the very late-type
spirals/irregulars at low redshift (Kennicutt 1992a,b) and even for the
nearby galaxies it is still  a matter of debate whether they are able
to sustain the current SFR for an extended period of time and whether
they are experiencing a continuous or an episodic SF history.  

In order to better understand the nature of the e(b) galaxies in our
sample, we have estimated their metallicities using the $R_{23}$ index
which is a ratio of line fluxes (Pagel et al.\ 1979; Edmunds \& Pagel
1984):  $R_{23}=({\rm[O{\sc ii}]3727+[O{\sc iii}]4959,5007})/\rm
H\beta$.  The measured abundances of the e(b) galaxies (Appendix~A) are
significantly lower than those of any other spectral class and are
comparable to those in low luminosity, very late type spirals and the
most luminous irregulars at low redshift.

\subsection{The e(c) class}

By definition, the e(c) spectra are similar to those of typical
present-day spirals: they are defined to have  weak to moderate [O{\sc
ii}]$\lambda$3727 emission ([O{\sc ii}] detected, but with EW([O{\sc
ii}])$<40$\AA) and weak to moderate $\rm H\delta$ absorption ($\rm
EW(H\delta) < 4$\AA).  This broad range in emission and absorption line
strength has been chosen to encompass the observed characteristics of
all the normal spirals of Sa type or later  in the sample of nearby
galaxies of Kennicutt (1992a,b).  The spiral--like model spectra used
in this paper (some of which are shown in the Fig.~2a) fall in this
spectral class. 

The modeling also shows that there is  an alternative way of
reproducing an e(c) spectrum: starbursts of sufficiently long duration
($>> 0.1$\,Gyr) can in their later phases mimic the e(c) classes.
After an initial e(b) phase with very strong emission lines lasting
$\sim 0.1$\,Gyr, the starbursting galaxy displays an e(c) spectrum  as
long as the star formation continues.  This is shown in Fig.~3, for a
burst model as in Fig.~2c, in which the burst lasts 1\,Gyr  and ends at
$\tau=0$ with no further SF.

In this model $\sim 20$\% of the galactic mass is turned into stars
during the burst, but the SFR in the  starburst phase (assumed to be
constant) is just  3 times the rate before the burst begins. Of course
at this level the use of the term `starburst' is arbitrary: for this
model the timescale for gas exhaustion is not a small fraction of the
Hubble time.

\hbox{~}
\vspace{-0.4in}
\centerline{\psfig{file=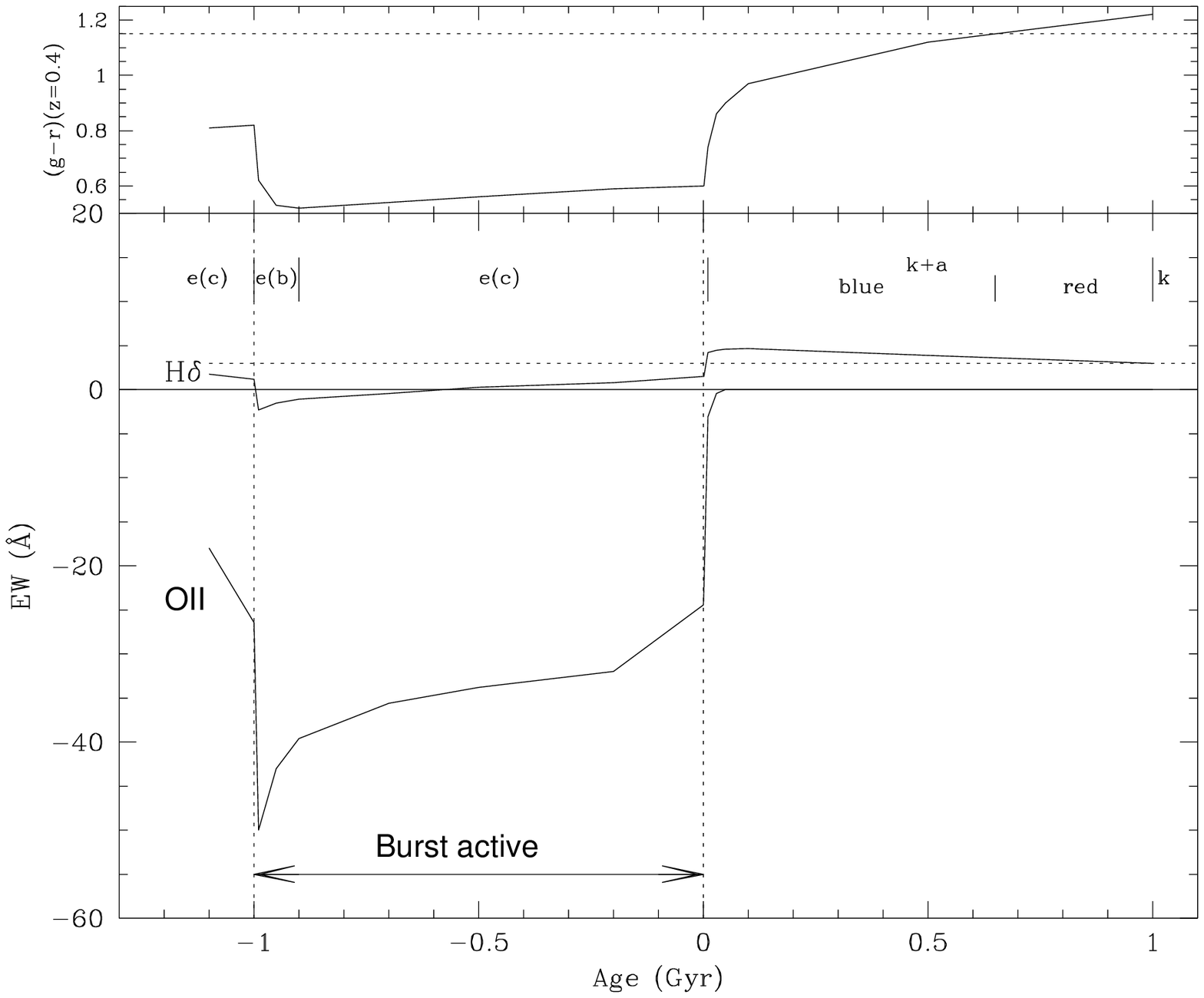,angle=0,width=3.5in}}
\vspace{-0.4in}
\noindent{\scriptsize
\addtolength{\baselineskip}{-3pt} 
\hspace*{0.3cm} Fig.~3.\ Modeling of a `long burst'. The plot shows the
evolution of  the [O{\sc ii}] and $\rm H\delta$ equivalent widths
during and after a burst lasting 1\,Gyr and ending at $\tau=0$. The two
dotted vertical lines delimit the period of the burst.  k+a model
spectra lie above the dotted horizontal line at  EW($\rm
H\delta$)=3\AA.  The spectral classification at each stage is shown in
the top part of the lower panel, including the distinction between blue
and red k+a phase.  For ages $< -1$\,Gyr a `spiral-like' model is
assumed (Fig.~2a).  The top panel shows the $(g-r)$ color if the galaxy
was observed at $z=0.4$ at any time. The dotted line in the top panel
represents the limit between red and blue galaxies.

\addtolength{\baselineskip}{3pt}
}

\subsection{The e(a) class}

The e(a) class includes all the galaxies with EW$(\rm H\delta)> 4 $\AA\
and measurable [O{\sc ii}] emission.  We have chosen to treat these as
a separate class from the e(c) spectra because their $\rm H\delta$ line
is too strong compared to those of normal low-redshift galaxies of any
Hubble type (Kennicutt 1992a).  Further suggestions of the unusual
nature of this class comes from our modelling of exponentially-decaying
SFR, with a range of timescales, which confirms that a continuous SF
history does not produce spectra with a net EW$(\rm H\delta)>4$\AA\
in absorption, when the emission filling is taken into account.  

In the D99 spectroscopic catalog there are 44 e(a) spectra out of a
total of 399 cluster members with  an assigned spectral class.  The
fraction of e(a) spectra in the magnitude limited sample is around
$\sim 10$\%.  Although spectra with these characteristics are present
in other spectroscopic surveys of distant clusters (e.g.\ CS87; Couch
et al.\ 1994, 1998; Fisher et al.\ 1998), their relevance and
interpretation  have not been extensively discussed in previous works.
We now present dust-free modeling of this spectral class, followed by a
discussion of the role of dust in defining the characteristics of
low-redshift galaxies with e(a) spectra.

Since the e(a) spectra display emission lines, only models with active
star formation at the time of observation have been considered; an
extensive region of parameter space and various star formation
histories have been investigated (Fig.~2).  The spectral
characteristics of e(a) galaxies can be reproduced assuming that, after
a recent strong starburst, the galaxy resumes a much lower SFR. Both of
these properties (recent burst and a low current SFR) are necessary to
explain the observed spectra.

Models with a regular, smooth SF history, with no burst and no
truncation, (Fig.~2a) never reach EW($\rm H\delta)>2$--3\AA, even for
the SEDs typical of later morphological types.  Truncated models
(Fig.~2b, solid line) cannot reproduce the [O{\sc ii}] line since they
lack any present star formation.

We also investigated a class of models in which a spiral-like SFR drops
to a much lower, but non-zero, level without experiencing a starburst
(dotted line, Fig.~2b). All the models of this class have $\rm
EW(H\delta)<4$\AA\ and therefore they are unable  to reproduce e(a)
spectra. 

Models with truncated SF at the end of the burst (Fig.~2c) also
experience a very short e(a) phase, however this phase is far too short
to account for the large number of e(a)'s observed and furthermore the
EW([O{\sc ii}]) of these models are generally weaker than the observed
values.

The only class of dust-free models that reproduce the strong $\rm
H\delta$ in absorption and the [O{\sc ii}] in emission are
post-starburst galaxies with  some residual SF (Fig.~2d). As in the
case of the k+a/a+k galaxies, the high  EW$(\rm H\delta)$ requires a
high fraction of the galactic mass to be involved in the burst
(typically 20\% or more for a standard IMF) and this result is
independent of the burst duration, as long as it is $< 1 $\,Gyr.  We
will see that this interpretation of the e(a) spectra predicts  a
bright phase with strong emission lines (e(b) spectrum) which  is not
observed.

Figure~4 shows a sequence of models at 0, 0.01, 0.1, 0.5, 1 and 2\,Gyr
after the end of a strong burst of SF (about 35\% in mass), lasting
0.1\,Gyr and superimposed on a spiral-like `Sc' SF history.  This is a
model of the type shown in Fig.~2d, in which at the end of the burst
the galaxy resumes the regular spiral-like SFR and during the burst the
SFR is 100 times the rate of the undisturbed galaxy.  The model starts
with an e(b) spectrum ($\tau=0$, where $\tau$  is the time elapsed
since the end of the burst) with strong emission lines including  $\rm
H\delta$ in emission; at $\tau=0.01$\,Gyr the emission has already
decreased significantly and the $\rm H\delta$ line is quickly moving
towards high EW in absorption (e(c)/e(a) spectrum).  The galaxy is an
e(a) at $\tau=0.1$\,Gyr and it remains in this spectral class until
$\tau=1$\,Gyr, when the EW($\rm H\delta$) in absorption begins to
decrease: from this moment on the galaxy has an e(c) spectrum as long
as there is on-going star formation.  This model remains blue
(according to the definition given in \S2) during the whole evolution.

\hbox{~}
\centerline{\hspace*{0.7cm}\psfig{file=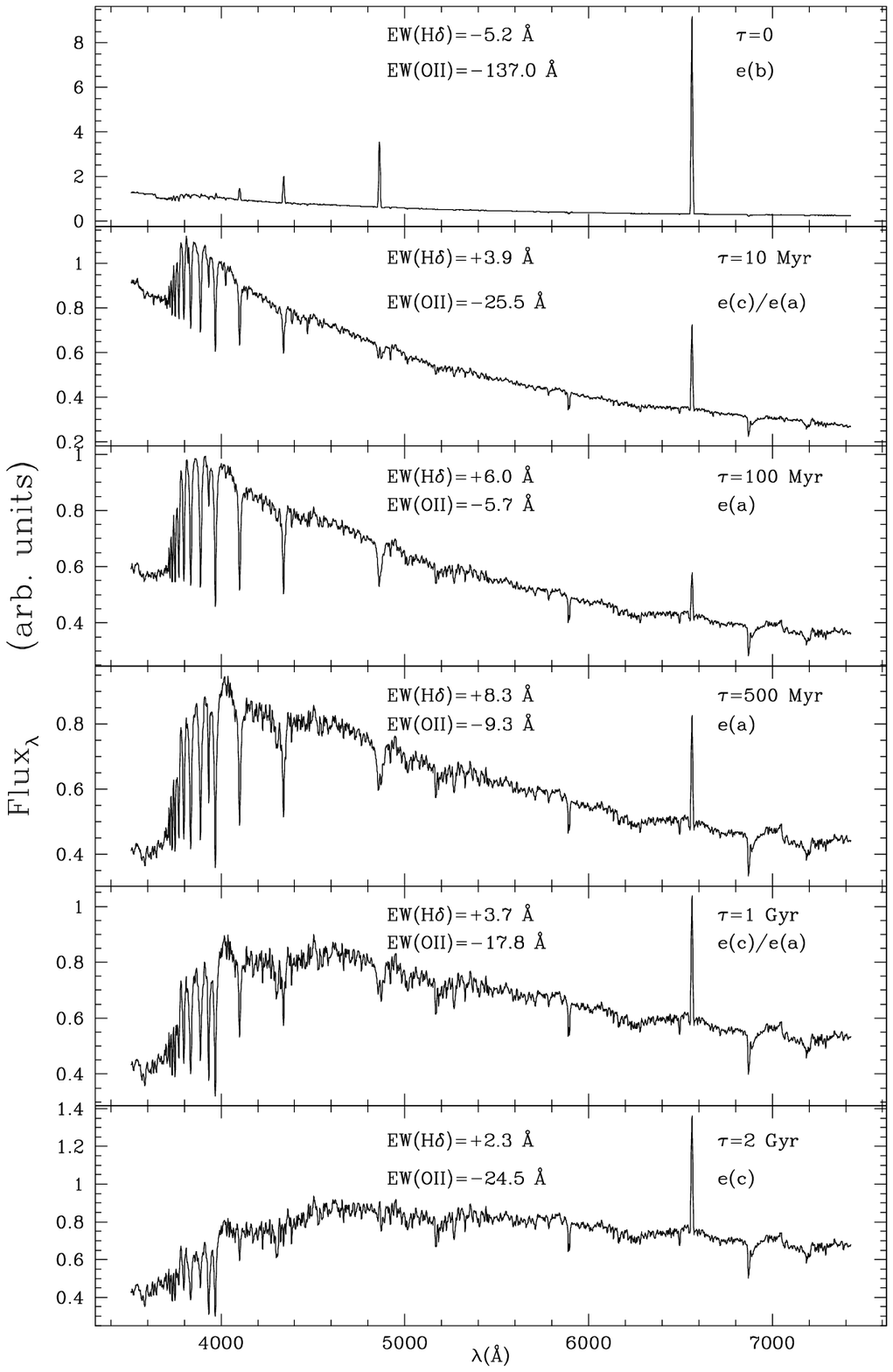,angle=0,width=4.2in}}

\noindent{\scriptsize
\addtolength{\baselineskip}{-3pt} 
\hspace*{0.3cm}
Fig.~4.\ Spectra of a burst model with residual SF (Fig.~2d).  $\tau$
is the time elapsed since the end of the burst.

\addtolength{\baselineskip}{3pt}
}

Our dust-free models therefore suggest that e(a) spectra can be
obtained whenever a strong burst of star formation is followed by
continuing star formation at a low level.  In this scenario e(a)
galaxies are post-starburst galaxies and they differ from  the k+a/a+k
population only by having some residual star formation.  We now discuss
the possibility that the e(a) galaxies themselves are in the strong
burst phase, heavily obscured by dust.

\subsection{The role of dust}

The e(a) class is, among our spectral classes, the one that has been
the least discussed and is probably also the least understood.  To gain
more physical insight into the nature of galaxies in the e(a) class we
have searched  for examples of this class of galaxy in the local
universe.  We find good examples of e(a) spectra in the spectral atlas
of merging galaxies by Liu \& Kennicutt (1995a LK95; 1995b), which is
composed of merging or strongly interacting systems, as judged by
either optical or near infrared imaging.  We reanalyzed their spectra
in the same manner as  D99 and  found that around 40\% of their
galaxies have an e(a) spectrum (16/39), while at most 7\% (1/14) of the
normal (non--merging, non--Seyfert) galaxies in Kennicutt (1992a,b)
show an e(a) spectrum.

The e(a) galaxies in the LK95 sample are known to be dusty, merging,
starburst galaxies: they are all strong FIR emitters and many of them
are classified as ULIRG (Ultraluminous IR Galaxies, $\log_{10} \,
L_{fir}> 11.5 \, \log_{10} L_{\odot}$ according to the LK95
definition).\footnote{The strongest ULIRG are believed to be often the
site of both powerful nuclear starbursts and AGNs, with the fraction of
AGNs increasing with the FIR luminosity, particularly at $L_{ir}>
10^{12.3} L_{\odot}$.  (Sanders \& Mirabel 1996; Duc et al.\ 1997;  Kim
et al.\ 1998). While the connection between nuclear starbursts and AGN
in ULIRGs is still under debate, it is well established that strong
starbursts occur in the nuclei of all the very luminous IR galaxies (Wu
et al.\ 1998 and references therein).}

A more detailed comparison of the similarities of the distant e(a)
class and the galaxies in the LK95 sample is possible using the ratio
of [O{\sc ii}] and $\rm H\alpha$ emission line strengths.  The e(a)
spectra, both in the local merging sample and in the distant clusters,
tend to have lower EW([O{\sc ii}])/EW($\rm H\alpha$ + [N{\sc ii}])
ratios than normal galaxies: in Fig.~5 they are represented  by the
filled symbols and they are found preferentially below the relation
EW([O{\sc ii}])= 0.4 EW($\rm H\alpha$ + [N{\sc ii}]) valid for the
normal galaxy sample of Kennicutt (1992a,b).  These low EW([O{\sc
ii}])/EW($\rm H\alpha$ + [N{\sc ii}]) ratios can be explained as an
effect of dust: the line emission comes from the H{\sc ii} regions,
where the young stars are still embedded in large amounts of
interstellar matter. In a very dusty environment, both lines suffer
strong extinction, [O{\sc ii}] being more affected than the $\rm
H\alpha$.\footnote{We note that there is the possibility that some of
the non--emission line spectra are actually e(a) galaxies with an
extremely low [O{\sc ii}] line, where the $\rm H\alpha$ line is out of
our spectrum, as it is observed in some cases in the UCM Survey
(Gallego, private communication).} However, the continuum underlying
the lines probably comes from stellar populations which are more widely
distributed throughout the galaxy, and hence suffers a lower extinction
than the lines. The net result is a lower ratio of the equivalent
widths of the two lines.  Another possible explanation for the low
[O{\sc ii}]/$\rm H\alpha$ ratios is high excitation, but this is ruled
out for our cluster sample of e(a)'s, which do not generally show
strong [O{\sc iii}]$\lambda$5007 lines.

\hbox{~}
\vspace{-0.2in}
\centerline{\psfig{file=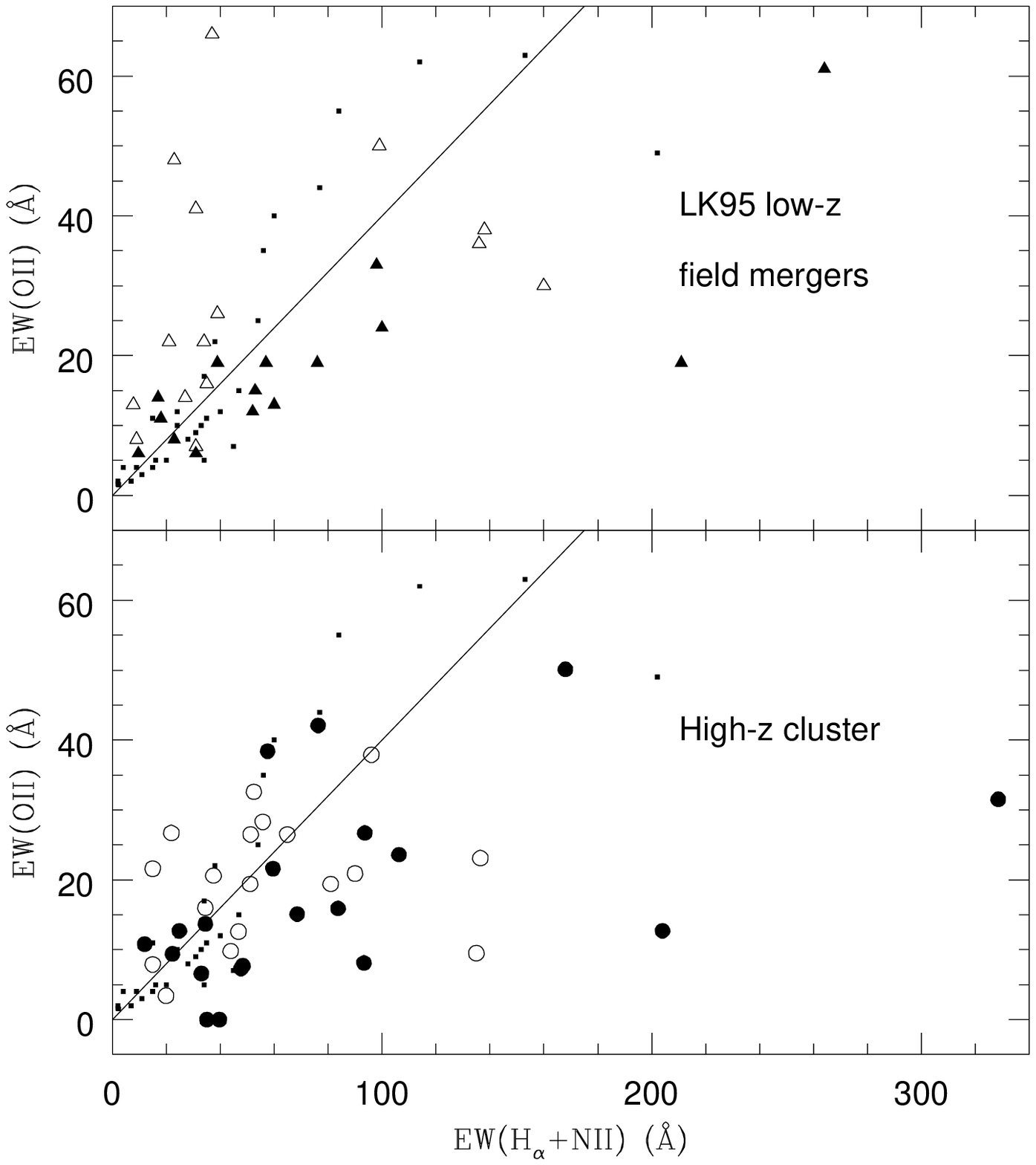,angle=0,width=4.2in}}
\vspace{-0.3in}
\noindent{\scriptsize
\addtolength{\baselineskip}{-3pt} 
\hspace*{0.3cm} 
Fig.~5.\ Comparison of the LK95 sample of merging galaxies with our
distant cluster sample. Filled symbols represent e(a) spectra. In the
top panel empty triangles represent merging galaxies of non-e(a)
spectral class, while in the lower panel the empty circles are cluster
members with e(c) spectra.  In the distant cluster members the measured
EW($\rm H\alpha$) has been multiplied by a factor 1.5 if the [N{\sc
ii}] line was not already included in the measurement, assuming [N{\sc
ii}]/$\rm H\alpha \simeq 0.5 $  on average (Kennicutt 1992b).  The
distant cluster e(c) galaxies are distributed equally about the fit for
normal galaxies in the local Universe (Kennicutt 1992a,b, small filled
rectangles), as expected if they are the counterparts  of normal local
spirals.  Both plots are uncorrected for extinction.

\addtolength{\baselineskip}{3pt}
}
\smallskip

We can see from Fig.~5 that using the EW($\rm H\alpha$) to estimate the
current SFR for e(a) galaxies instead of EW([O{\sc ii}]), would tend to
increase the apparent SFR by factors up to $\gs 2$--3.   This revision
means that we should interpret the e(a) population as dusty starbursts
\footnote{We note in passing that at least one e(a) cluster galaxy,
\#834 in Cl\,0024$+$16, has been detected at sub-mm wavelengths
indicating a substantial dust content (Smail et al.\ 1998)},\footnote{
Analysis of 123 spectra of Very Luminous IR Galaxies (Poggianti \& Wu,
in preparation) from the sample of Wu et al.\ (1998) confirms the
exceptionally high fraction (about 50\%) of e(a) spectra among
FIR--luminous galaxies.}\ and that the SFR and histories derived
for e(a) spectra from dust--free models (\S3.4) are misleading. In
contrast, the spectral properties and [O{\sc ii}]/$\rm H\alpha$ ratios
of the e(c) and the e(b) spectra don't show evidence for a high dust
extinction and the dust--free models discussed in \S3.2 and \S3.3 can
be considered appropriate.

In a dusty starburst a high fraction of the bolometric luminosity is
emitted at FIR wavelengths and since this is an optically--selected
sample, one might worry about biases in the selection of the e(a)
sample relative to the other (less reddened) spectral classes.

The competing effects of dust and starbursts on the luminosities of the
e(a) are complicated to investigate. The net effect depends upon the
details of the mixing of the dust with the young and old stars within
the galaxy. For likely burst strengths we would expect a brightening
during the peak of the burst phase of 1--2 magnitudes in the restframe
$V$-band (Fig.~6).  However, the observed ratio of H\,$\alpha$ to
[O{\sc ii}] equivalent widths suggests that the visual extinction to
the emission line regions is in the range 1.2--2.5
magnitudes.\footnote{This is found adopting as {\it attenuation curve}
the standard {\it extinction curve} of the diffuse medium in the
Galaxy. We are forced to this arbitrary assumption by the lack of
alternatives.} As the majority of the blue and visual luminosity
resulting from the starburst is also likely to reside in these regions
we expect that the two effects will roughly cancel.  Only after the
stars have diffused out of these high extinction regions will the dust
become less effective, but the burst population will also have evolved
and faded by this time.

This cancellation suggests that there will be a relatively modest
variation in optical luminosity for different burst strengths. Using
the far-infrared luminosity as a tracer of the strength of the
starburst we see exactly this behaviour at low-$z$.  In fact, a  range
of over 3 orders of magnitude in $L_{FIR}$ for infrared--selected
galaxies at low--redshift is accompanied by less than a factor of 3--4
change in the optical luminosity (Sanders \& Mirabel 1996).  As an
extreme case, the Ultraluminous Infrared Galaxies ($L_{FIR}> 10^{12}
L_{\odot}$) have optical luminosities comparable to present--day
$L_{\ast}$ (Sanders \& Mirabel 1996 and references therein).  The FIR
properties of our e(a) galaxies are probably less extreme, but even so
their optical luminosities are not expected to be less than 1/2--1/3 of
$L^\ast$. The D99 spectral catalog provides a representative luminosity
distribution of cluster members at $M_V < -19 + log_{10} h$, 1.2
magnitudes below the $M^\ast_V$ of spiral galaxies at the cluster
redshift (S97).  We conclude that in general we expect the optical
luminosities of the e(a) population to be crudely comparable to that
expected for a non-bursting system.  This suggests that there is no
strong bias in our characterisation of this population due to its dusty
nature.

It is clear that galaxies with e(a) spectra appear in the low--redshift
universe as well (see also Carter et al.\ (1988); Fig.~1 from Zabludoff
et al.\ (1996); Gallego et al.\ (1997)).  Their frequency in the field at
low-$z$ as estimated from the Las Campanas Redshift Survey (8\%,
Hashimoto 1998) appears lower than in the distant field and clusters,
although this comparison is an uncertain one because of the difference
in the samples and the classification criteria.

At redshifts comparable to the D99 catalog, a population of Balmer
strong galaxies  with emission is present in  the field sample of the
Canada-France  Redshift Survey (see Fig.~15 in Hammer et al.\ 1997 and
Hammer \& Flores 1998).  Although a direct comparison cannot be made,
their Balmer index versus D4000 plot is equivalent  to our Fig.~1 and
the e(a) population is evident both at $0.5 < z < 0.7$ and at $0.7 < z
< 1$. This is in agreement with the results of the field sample in D99
(see \S5.2 and Fig.~1).

We will now try to identify the possible evolutionary paths among the
spectral classes and hence discuss how the interpretation of e(a)
spectra as dusty starbursts suggests they are good candidates for
precursors of the k+a/a+k population.

\section{Predicted evolution of the spectral classes}

We have seen in the previous section that the high proportion of
k+a/a+k spectra observed in distant clusters indicates that a moderate
fraction of cluster members were forming stars in the recent past but
that this activity terminated at an epoch close to the time at which
they are observed.  In this sense their spectral properties indicate a
sharp decrease in star formation activity.  At the same time, in many
cases the strong $\rm H\delta$ observed requires a burst of star
formation prior to the sharp decline.  Given that  the post--starburst
models spend the second half of their k+a/a+k lifetime having more
moderate $\rm H \delta$ strengths which progressively decline  towards
the 3\AA\ threshold (Poggianti \& Barbaro 1996), at least some of
the k+a/a+k spectra in the range $3< EW(\rm H \delta) < 5 $ \AA\
are expected to be the remnants of much stronger cases.  Therefore, on
the basis of standard--IMF models, we find that a large fraction,
possibly all,  of the k+a/a+k galaxies in our sample are likely to have
experienced a recent starburst. Although this fraction is surely
significant, a quantitative estimate is problematic, given the
uncertainty in the exact threshold between post--starburst and
truncated models and the modest incompleteness of the sample for
EW($\rm H\delta$)  between 3 and 5\AA\ (D99).  

The post-starburst interpretation of the k+a/a+k spectra necessarily
poses the question of where are the starbursts, i.e.\ what is the
progenitor population of the post-starburst galaxies?  Our modeling
shows that if no dust extinction is involved and a standard IMF is
assumed, then the combination of a high SFR (required from the Balmer
lines of the post-starburst galaxies) and short timescales (typical of
starbursts) inevitably leads to emission lines with high equivalent
widths.  About 5\% of our spectra show such strong emission lines (e(b)
class), and therefore at face value these would be the most obvious
candidate progenitors of the strong post-starburst k+a/a+k population.

We now show that the evolutionary scenario from e(b) to k+a/a+k spectra
fails to account for the relative luminosity distributions observed for
these two classes in the distant clusters.  As discussed in D99,
galaxies with e(b) spectra are conspicuously fainter than all the other
spectral classes. Their typical luminosity range is also shown on the
left side of Fig.~6, which presents the expected luminosity evolution
of a galaxy experiencing a burst and then no subsequent star
formation.  Assuming the burst lasts 0.1\,Gyr and ends at $\tau=0$ in
the plot -- corresponding to $z=0.4$ -- the absolute $V$ magnitude
evolves as shown for each of three different values of the galactic
mass fraction of the burst ($\Delta g$).

Even for a modest $\Delta g$ value, when the SF ceases, the model has
already faded by 0.4--0.6 mag (depending on the intensity of the burst)
at $\tau =0.1$\,Gyr, in  comparison with the average magnitude during
the e(b) phase.  By the time the galaxy ends its k+a phase, it has
faded by 1--1.5 mag.
  
We conclude that the bulk of the e(b) galaxies are not the progenitors
of the k+a/a+k class we observed and that they do not evolve into any
other spectral class visible in our sample:  our analysis of their
metallicities confirms that the properties of the e(b) galaxies
resemble those of late-type, low-luminosity galaxies at low redshifts.
If their destiny is to remain totally passive  in their subsequent
evolution, their spectral properties and magnitudes by $z=0$ will be
similar to those of passive dwarfs, as suggested by Koo et al.\ (1997)
(see also Wilson et al.\ 1997 and Couch et al.\ 1998).

\hbox{~}
\vspace*{-0.5in}
\centerline{\psfig{file=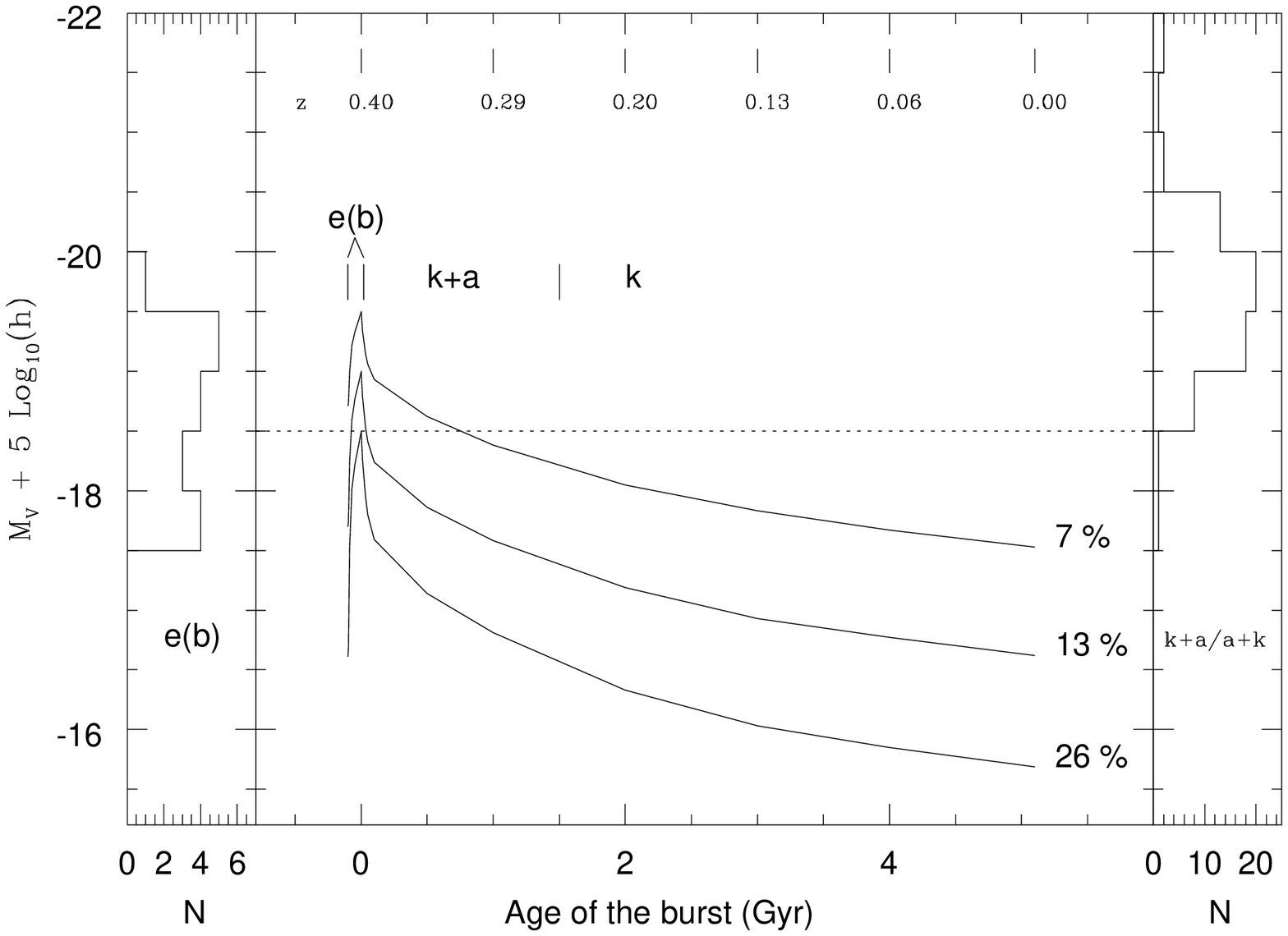,angle=0,width=3.7in}}
\vspace{-0.3in}
\noindent{\scriptsize
\addtolength{\baselineskip}{-3pt} 
\hspace*{0.3cm}
Fig.~6.\ The evolution of the absolute $V$ magnitude of three model
galaxies which suffer a star burst and no subsequent star formation.
The galactic mass fractions involved in the burst are shown on the
right side of the plot; the flux from any emission line lying within
the $V$ band is not included.  Lines start at the beginning of the
burst, which lasts $0.1$\,Gyr and ends at $\tau=0$ taken to be
$z=0.4$.  Prior to the burst a `late spiral-like' star formation
history is assumed (Fig.~2c); this is a conservative choice that
minimizes the luminosity evolution.  The three models have been
displaced arbitrarily at intervals of 0.5 mag at $\tau=0$ for display
purposes.  All of these models exhibit an e(b) spectrum during the
burst, followed by a k+a phase lasting about 1.5\,Gyr and finally
evolve into k galaxies.  The dotted line shows the magnitude limit for
k+a/a+k detection. The relation $z$--age is given for $h=0.5$.

\addtolength{\baselineskip}{3pt}
}

In Section 3.3 we have seen that, in principle at least, some of our
observed e(c) galaxies could be `long starbursts' and evolve into
k+a/a+k galaxies  later on, but it is more likely these are simply
galaxies which are forming stars in a regular, continuous manner with a
spiral--like star formation history.

The question of whether an e(c) galaxy could evolve into a k+a {\it
without} experiencing a starburst phase has been addressed in previous
works  (CS87; Newberry et al.\ 1990; Poggianti \& Barbaro 1996). As
discussed in \S3.1, models with a truncated star formation do evolve
from e(c) to k+a, but can only account for the weakest k+a types ($\rm
EW(H\delta)< 5$\AA), or at most half of our sample of k+a/a+k
galaxies.   

Another class of emission line galaxies, e(a), is a better candidate
for the progenitors of the k+a/a+k class.  On the basis of dust-free
models, the e(a) spectra appear associated with {\it post-starburst}
galaxies with some residual star formation.  If the SF eventually
ceases, the spectrum evolves from the e(a) into  the k+a/a+k class.
The evolutionary sequence of this class of models starts presumably
from an e(c) spectrum and ends up in a k+a/a+k or e(c) galaxy depending
on whether the SF comes to an end:

\begin{center}
\begin{picture}(250,100)\thicklines
  \put(0,50){e(c) (?)}
  \put(40,50){$\Rightarrow$}
  \put(60,50){e(b)}
  \put(80,50){$\Rightarrow$}
  \put(100,50){e(a)}
  \put(120,60){$\nearrow$}
  \put(130,70){e(c)}
  \put(120,40){$\searrow$}
  \put(130,30){k+a/a+k}
  \put(175,30){$\Rightarrow$}
  \put(195,30){k}
\end{picture}
\end{center}

Although this evolutionary scenario is capable of interpreting an e(a)
spectrum, this model implies a bright phase with strong emission lines
(e(b) spectrum) which is not observed.  In this model e(a) and k+a/a+k
galaxies are post-starburst galaxies and the only difference between
these two classes is a low level of SF in the e(a) class. 

However, the striking resemblance of our distant e(a) class with the
spectra of low-$z$ dusty starbursts, and the analysis of their spectral
properties, in particular, their low EW([O{\sc ii}])/EW($\rm H\alpha$)
ratios, indicates  that the e(a) galaxies are in fact starbursts which
contain substantial amounts of dust and that 
dust-free models are unsuitable to estimate their star formation rates
and histories. 

\hbox{~}
\vspace*{-0.5cm}
\centerline{\hspace{1.0in}\psfig{file=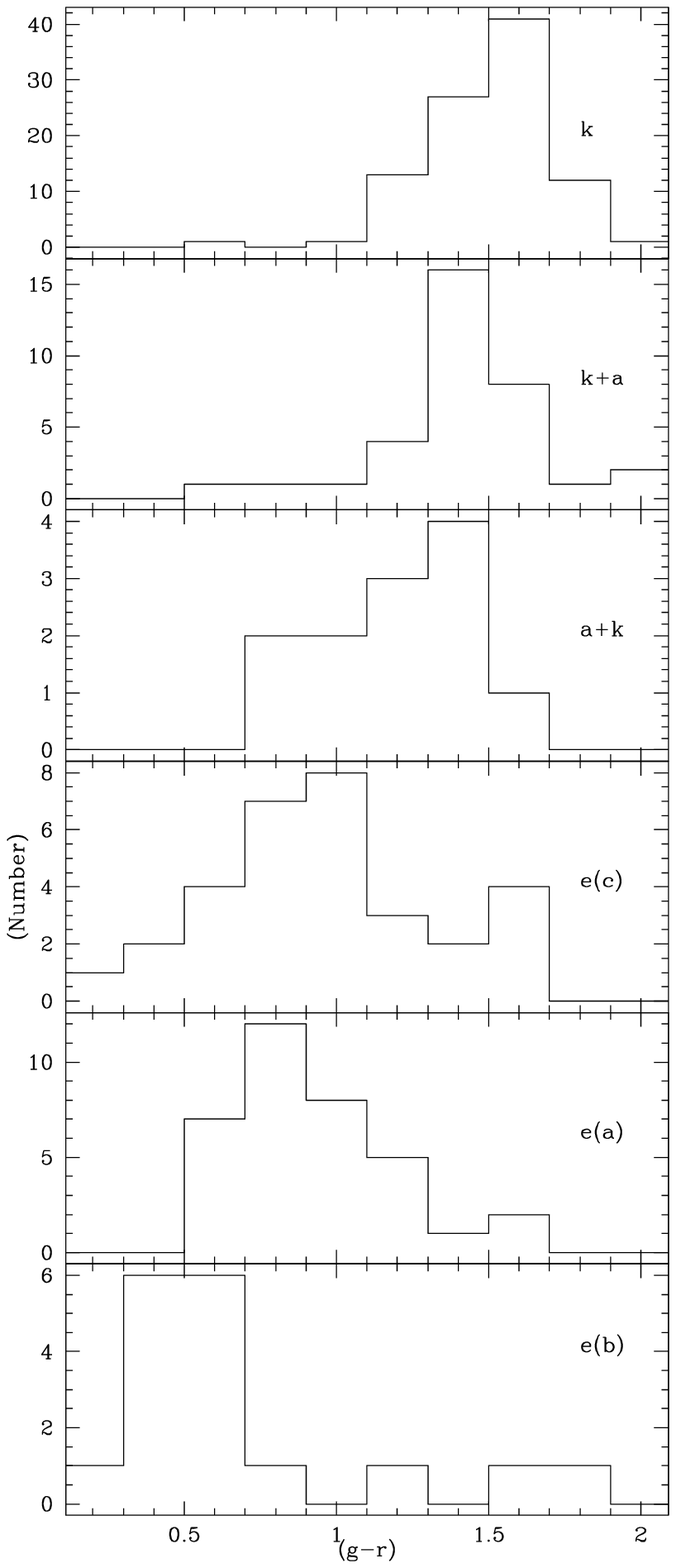,angle=0,width=4.2in}}

\noindent{\scriptsize
\addtolength{\baselineskip}{-3pt} 
\hspace*{0.3cm}
Fig.~7.\ Color distribution as a function of the spectral class.
All colors are K--corrected to $z=0.4$.

\addtolength{\baselineskip}{3pt}
}
\smallskip

We suggest that the e(a)'s, rather than the very rare bright e(b)'s,
may be the missing link between the continuous star forming galaxies
and the post--starburst k+a/a+k class so abundant in the distant
clusters.  The e(a)'s and k+a/a+k's would then be galaxies in two
distinct evolutionary phases, starburst and post--starburst
respectively.  Identifying a strong $\rm H \delta$ line in
emission--line spectra then represents a very powerful method to
determine the incidence of a population of dusty star--forming galaxies
up to high redshift and recognize the cases where the optically--based
SFR estimates are unreliable.

Furthermore, if the e(a) galaxies are dusty progenitors of the k+a/a+k
galaxies, then it suggests that dust reddening might also affect the
observed properties of some of the k+a/a+k galaxies.  In particular
dust reddening may also be responsible for the class of very red k+a
galaxies mentioned in \S3.1.  As discussed there, these galaxies have
broad-band colors which are too red given the strength of the
post-starburst features seen in the spectra.  An internal reddening of
$A_V=0.5$ mag  was found sufficient by CS87 to reproduce the reddest
colors of the H\,$\delta$-strong population.  

The absolute magnitude and color distributions (Fig.~7) as a function
of the spectral class are consistent with the scenario outlined above.
In fact the great majority of e(b)'s are very blue, most of the e(a)
and e(c) types and a significant fraction of a+k are blue, while the
great majority of k+a and k types are red.  Fig.~7 also shows that most
of the blue galaxies responsible for the Butcher--Oemler effect are
star-forming galaxies, since $\sim 85$\% of them have emission lines:
13\% e(b),  41\% e(a), 31\% e(c), 6\% a+k, 6\% k+a, 4\% k. The
original Butcher--Oemler magnitude and radius cutoffs have not been
applied to derive these fractions. If the absolute magnitude cutoff is
applied the contribution of e(b) galaxies is drastically reduced.
  
To understand the environmental processes responsible for the
post--starburst population and the connection between the
spectrophotometric and the morphological evolution, we next discuss the
properties of the different cluster populations in relation to their
environments and morphologies from our {\it HST} imaging.  Our aim is
to search for ways to distinguish the underlying physical mechanisms at
work in the formation of the k+a/a+k class, the k--type spiral galaxies
and more generally the S0 population (Dressler et al.\ 1997; van Dokkum
et al.\ 1998).   

\section{Environmental and morphological trends in SF activity}

The basic relation between star formation and morphology for cluster
galaxies in our sample has been presented in Dressler et al.\ (1998,
their Fig.~7). They find that the passive k--type population is
dominated by early-type galaxies, E and S0, but that it also contains a
large number of later-type spirals, stretching out to Sd/Irr.  The
active cluster populations are typically populated by intermediate and
late-type spirals: e(a) and e(c) are predominantly Sa--Sd/Irr galaxies,
while e(b)'s are mostly Sc--Sd/Irr.  The morphological distributions of
the post-starburst galaxies lie intermediate between the passive and
active cluster populations:  the a+k distribution is similar to that of
the e(a)/e(c) galaxies, while the k+a distribution is relatively flat
in types from E to Sc and therefore the k+a/a+k spectra mostly belong
to spirals (as was found by Dressler et al.\ 1994; Wirth et al.\ 1994;
Couch et al.\ 1998).  

Looking at the distribution of spectral classes within each Hubble
type, ellipticals, S0 and Sa galaxies have predominantly passive
k--type spectra, with a tail of k+a/a+k and emission line classes. Sb,
Sc and Sd/Irr galaxies show in most cases emission lines, but still
have an important fraction of k+a/a+k and k spectra which, although it
decreases for later Hubble types, is still seen even in Sd/Irr galaxies.
It is remarkable that only a small fraction (about 10\%) 
of the spirals of types Sa--Sc in the core regions of the clusters
have the spectral characteristics which are common in low--redshift
normal spirals (e(c) spectrum). The great majority are either 
`too active' (e(a) and e(b) classes, 22\%) or `too inactive' 
(k+a/a+k and k classes, 25 and 43\% respectively).

\subsection{Field versus cluster galaxy properties}

The great advantage of the spectral catalog in D99 is that it provides
a direct comparison of the field and cluster populations at
intermediate redshifts, which enables us to search for environmental
variations in the properties of galaxies.  The fraction of galaxies as
a function of the spectral class are given for each cluster and for the
total cluster and field samples in Table~4.  They have been corrected
for the morphological selection and they are given as fractions of the
total number of galaxies with assigned spectral class, excluding the
e(n) class. 

In contrast to the cluster environment, only a small proportion of the
field spiral population falls into the k--type class, and these are
typically early-type spirals.   In the whole field sample more than
half of the Sa--Sc spirals have an e(c) spectrum (54\%), while the
proportion of `too inactive' spirals is much smaller (5\% k+a/a+k, 16\%
k), and the remaining 24\% are e(a)'s or e(b)'s.  

Although the distant clusters included in our sample do contain
substantial populations of star forming galaxies, it is also true that
these emission-line galaxies are less common in the clusters than in
the surrounding field (see Table~4). Moreover, at a fixed Hubble type,
the frequency of galaxies with low or no measurable [O{\sc ii}]
emission  is significantly higher in the clusters (D99).  If the
cluster triggers the starbursts responsible for the strong k+a/a+k
cases, it must do this without significantly enhancing the [O{\sc ii}]
emission (D99, c.f.\ Balogh et al.\ 1997, 1998).  This is conceivable,
since the [O{\sc ii}] line is not a reliable indicator of current star
formation in  dusty galaxies (such as the e(a) galaxies in our
interpretation), and the fact that the e(a) fraction in the cluster,
although diluted by the passive pre--existing population, is similar to
the one in the field seems to suggest that the cluster environment is
responsible for triggering some of the dusty starbursts.  A definitive
conclusion cannot be reached on the basis of the present sample due to
the large statistical errorbars on the observed fractions: comparing
the field and cluster proportion of e(a) spectra in the supposely
recently infallen population (Sa to Sd/Irr), there is only weak (1
sigma) evidence for a higher incidence of dusty starbursts in the
clusters.  Larger samples will be needed to assess the issue.

It is clear from the discussion above (see also Fig.~1) and in D99 that
the most striking feature of the galaxy populations in the cores of
distant clusters, especially when compared to the field at the same
epoch, is not so much the presence of star forming galaxies in this
environment, but rather the large population of galaxies which exhibit
post-starburst features in their spectra: the k+a/a+k class.  The high
fraction of post-starburst galaxies seen in our cluster sample is not
mirrored in the surrounding field population and argues for an
environmental mechanism for either the formation or prolonged
visibility of this phase (D99).  
 
\begin{table*}
{\scriptsize
\begin{center}
\centerline{\sc Table 4}
\vspace{0.1cm}
\centerline{\sc Fraction of galaxies as a function of spectral class}
\vspace{0.3cm}
\begin{tabular}{lccccccr}
\hline\hline
\noalign{\smallskip}
 {Cluster} & k & k+a/a+k & e(a) & e(c) & e(b) & e & N \cr
\hline
\noalign{\medskip}
Cl\,1447$+$23 & 0.38$\pm$0.13 & 0.06$\pm$0.05 & 0.26$\pm$0.10 & 0.26$\pm$0.10 & 0.04$\pm$0.04  & 0 & 21\cr
Cl\,0024$+$16 & 0.50$\pm$0.07 & 0.14$\pm$0.03 & 0.11$\pm$0.03 & 0.18$\pm$0.04 & 0.05$\pm$0.02 & 0.02$\pm$0.01 & 107 \cr
Cl\,0939$+47$ & 0.46$\pm$0.08 & 0.27$\pm$0.06 & 0.09$\pm$0.04 & 0.13$\pm$0.04 & 0.03$\pm$0.02  & 0.01$\pm$0.01 & 71\cr
Cl\,0303$+17$ & 0.29$\pm$0.07 & 0.16$\pm$0.05 & 0.11$\pm$0.05 & 0.23$\pm$0.07 & 0.11$\pm$0.05 & 0.10$\pm$0.04 & 51 \cr
3C\,295       & 0.53$\pm$0.14 & 0.28$\pm$0.10 & 0.07$\pm$0.05 & 0.07$\pm$0.05 & 0 & 0.04$\pm$0.04 & 25 \cr
Cl\,1601$+$42 & 0.59$\pm$0.10 & 0.26$\pm$0.07 & 0.07$\pm$0.03 & 0.05$\pm$0.03 & 0.03$\pm$0.02 & 0 & 58 \cr
Cl\,0016$+$16 & 0.55$\pm$0.12 & 0.32$\pm$0.09 & 0.08$\pm$0.04 & 0.03$\pm$0.03 & 0.03$\pm$0.03 & 0 & 29\cr
\noalign{\smallskip}
Total cluster      & 0.48$\pm$0.04 & 0.21$\pm$0.02 & 0.11$\pm$0.02 & 0.14$\pm$0.02 & 0.05$\pm$0.01 & 0.02$\pm$0.01 & 390 \cr
\noalign{\smallskip}
Field $0.35<z<0.60$ & 0.30$\pm$0.07 & 0.06$\pm$0.03 & 0.11$\pm$0.04 &
0.29$\pm$0.06 & 0.16$\pm$0.05 & 0.09$\pm$0.03 & 71 \cr
\noalign{\smallskip}
\noalign{\hrule}
\noalign{\smallskip}
\end{tabular}
\end{center}
}
\vspace*{-0.8cm}
\end{table*}

The differences between cluster and field populations and the dynamic
and spatial distributions of the `active' population (k+a/a+k and all
emission classes) are consistent with these being recently acquired
through infall from the field onto the clusters (D99), which in this
scenario are responsible for transforming emission-line galaxies into
k+a/a+k galaxies.  If the `active' population is due to infall from the
field,  then given that the timescale of the k+a/a+k phase is about
1\,Gyr, we can estimate the typical `survival time' of star formation
as the ratio between the number of galaxies with emission lines and the
number of k+a/a+k's. This timescale, ${\tau}_{em}$, is about 1.5\,Gyr
and corresponds to the average time elapsed between the moment these
galaxies `enter' the cluster (i.e.\ would be classified cluster members
according to our criteria) and the moment they stop forming stars and
therefore lose their emission lines. This corresponds to an average
`infall rate' of 8--12 galaxies per Gyr per cluster, depending whether
the k+a/a+k timescale is taken to be 1 or 1.5 Gyr.  Assuming that the
k+a/a+k population fades by a further magnitude before entering the k
class, this corresponds to about 12\% of the cluster core luminosity
acquired in the last Gyr.\footnote{We stress that this infall rate
depends on the membership criteria  adopted, which were chosen to
minimize the field contamination and to ensure that any galaxy in the
large--scale structure surrounding the clusters is retained (D99).}
Thus if, as expected in hierarchical models, the cluster has grown
substantially in the recent past this accretion has not consisted
solely of star-forming field galaxies, but is more likely to have been
a mix of passive and star-forming galaxies.

If the presence of these star--forming galaxies in the clusters is
related to the accretion of field galaxies, then their incidence is
expected to depend on the infall rate (Bower 1991; Kauffman 1995a,b;
Tormen 1998) and on the star--forming properties of galaxies in the
field, both of which are likely to vary with redshift.  The field
population itself shows significant evolution with a large proportion
of high--$z$ field galaxies actively forming stars.  In this scenario the
Butcher--Oemler effect, defined as the {\it disappearance} of these
star--forming galaxies at lower redshifts, is related to the cluster
environment and its capability to halt star formation, which amplifies
the trend of declining star formation at lower redshifts as compared to
the field.

\subsection{Morphological properties}

\subsubsection{Morphology of the e(a) population}

In the previous section we noted that a common feature of the e(a)
galaxies at low redshift is the evidence of mergers and/or strong
interactions.  ~From the morphological catalog (S97) it is possible to
look for similar effects through what we have called the disturbance
index ($D$) defined as : 0 -- little or no asymmetry, 1 or 2 --
moderate or strong asymmetry, 3 or 4 -- moderate or  strong
distortion.  The mean and median disturbance indices of each spectral
class are shown in Table~5. The errors quoted are bootstrap estimates
of the variances. The e(b) and the e(a) distributions are significantly
different from the k class and appear to be the most disturbed, but a
${\chi}^2$ test shows that the difference in the mean $D$ value between
the e(a) and the e(c) class is not highly significant (50\%).

\begin{table*}
{\scriptsize
\begin{center}
\centerline{\sc Table 5}
\vspace{0.1cm}
\centerline{\sc Mean and median Disturbance Index as a 
function of the spectral class}
\vspace{0.3cm}
\begin{tabular}{lccc}
\hline
\noalign{\smallskip}
Type & mean D & median D & M/T/I/C/null \\
\hline
\noalign{\smallskip}
k & 0.36$\pm$0.07 & 0.0$\pm$0.0 & 1/7/9/0/47 \\
k+a/a+k (red) & 0.84$\pm$0.22 & 1.0$\pm$0.5 & 2/2/4/0/14 \\
e(c) & 1.00$\pm$0.29 & 1.0$\pm$0.6 & 0/1/4/1/6 \\
k+a/a+k (blue) & 1.40$\pm$0.39 & 1.0$\pm$0.7 & 3/1/1/0/2 \\
e(a) & 1.47$\pm$0.18 & 2.0$\pm$0.5 &  4/4/0/0/12 \\
e(b) & 2.44$\pm$0.29 & 2.0$\pm$0.5 & 2/1/4/1/2 \\
\noalign{\smallskip}
\noalign{\hrule}
\noalign{\smallskip}
\end{tabular}
\end{center}
}
\vspace*{-0.8cm}
\end{table*}

We can also ask what the nature of the disturbance is in the e(a) and
e(c) classes, using the visual interpretations of the nature of
disturbance given in S97.   These interpretations are based on whether
the galaxy is an obvious merger (M), is undergoing a strong tidal
interaction (T), a weaker tidal interaction (I) or has  a chaotic
appearance (C). We list in Table~5 the number of galaxies belonging to
the M/T/I/C/null classes, where ``null'' means that no interpretation
is listed in S97.

Using these classes we see a strong difference between the e(a) and
e(c) populations, with all 8 of the e(a) galaxies for which an
interpretative class is listed being either mergers (4) or strong
interactions (4), compared to only one of the six e(c)'s which are
commented upon (the remaining five e(c) galaxies show either weaker
tidal features or a chaotic structure).

For the remaining e(a)'s and  e(c)'s either no disturbance was found or
no interpretation was given. Thus we suggest that at least a fraction
of the e(a) galaxies are involved in a merger or strong tidal
encounter. However, we also caution that it is difficult to assess the
significance of their disturbance in a sample where large disk galaxies
show in general a high incidence of non-axisymmetric structure in
comparison with their low-$z$ counterparts (S97).  To summarize,
although e(a) galaxies are indistinguishable from e(c) galaxies in
terms of frequency of disturbance compared to standard morphological
forms, e(a)'s appear much more likely to be connected to mergers or
strong interactions.

\subsubsection{Passive spirals}

At low redshift there are numerous examples of spirals with weak or
absent star formation, especially in clusters (see Koopmann \& Kenney,
1998, for a critical review of the correlation Hubble type to star
formation activity in clusters; Kennicutt 1998 with references
therein).  van den Bergh (1976) identified a class of `anemic' spirals
having smooth arms that lack bright stars and H{\sc ii} regions that
occur most frequently in clusters.  These galaxies are deficient in
neutral hydrogen but have normal CO emission (Kenney \& Young 1986,
1989; van den Bergh 1991).  The photometric properties of our k--type
spirals are similar to those of the very red (yet HI rich) spirals of
types Sb and Sc from the sample of Schommer \& Bothun (1983).  These
are luminous cluster spirals, but there are some field galaxies which
share the same properties (NGC1079, M31, see Schommer \& Bothun for
references), and  it is not well established whether this phenomenon is
more relevant in clusters.  In the Schommer \& Bothun sample some red
spirals exhibit an anemic appearance, while others have a normal spiral
structure with well--defined arms. 

A systematic study of the kind carried out in D99 has not yet been
undertaken for low--redshift clusters and therefore, although it is
known that passive spirals are present in clusters both at low and high
redshift, we cannot establish if they are more/equally/less prevalent
in low--$z$ clusters than in distant ones.  ~From our sample it is clear
that at $z \sim 0.4$ the passive spirals are much more common in the
clusters than in the field.  In Fig.~8 we compare the color
distribution of the k--type cluster spirals with the distributions of
other k--type Hubble types and the spirals with emission lines. k--type
spirals are \it significantly redder \rm than the spirals with emission
of similar Hubble type, and their color distribution is in fact
intermediate between the E/S0's distribution and the spirals with
emission. This confirms the reality of a separate population of
`passive spirals' which cannot be due to `slit effects' (the slit
missing a significant fraction of the disk light).  Note that the
difference in the color distribution between spirals with and without
emission lines is not due to a higher fraction of early Sa galaxies, as
this difference is still evident if one compares the distributions in
the two bottom panels of Fig.~8 separately for the Sa and the Sb--later
types.  We note that at present there is no evidence for a difference
in the radial distributions of passive (k) and emission--line spirals,
expected from the infall scenario, within the limited region imaged
with {\it HST}.

It is reasonable to envisage the existence of an evolutionary
connection between the k--type spirals and at least a fraction of the
population of k+a/a+k galaxies with spiral morphologies, since it is
obvious that the k--type spirals were forming stars at some point, and
therefore they must have experienced a k+a/a+k phase during their
evolution, soon after they stopped forming stars.  However, the
comparison of the number of k--type spirals (about 40 galaxies) with
the number of k+a/a+k spirals (27 galaxies) suggests that -- in the
assumption of a k+a/a+k production constant in time -- if the k+a/a+k
spirals evolve into k--types without changing morphology, they cannot
retain their morphological characteristics for a time much longer than
the k+a timescale, 1--1.5\,Gyr.  
  
The classification of a large proportion of the cluster spiral galaxies
as k and k+a/a+k type suggests that the process responsible for halting
star formation in these galaxies does not profoundly affect their
morphology, at least on a timescale comparable to the disappearance of
the Balmer--strong signature. The prevalence of k and k+a/a+k spirals
in clusters points to an environmental process capable of stopping star
formation (remove the gas reservoir) on a sufficiently short timescale
to allow strong Balmer features to be visible for quite a long period.
Moreover, the cessation of star formation must be achieved while
retaining the broad morphological characteristics of the galaxy. The
only process that the authors know of that acts exclusively on the gas
content of the galaxy is stripping due to the intracluster medium.

\hbox{~}
\centerline{\hspace{1.0in}\psfig{file=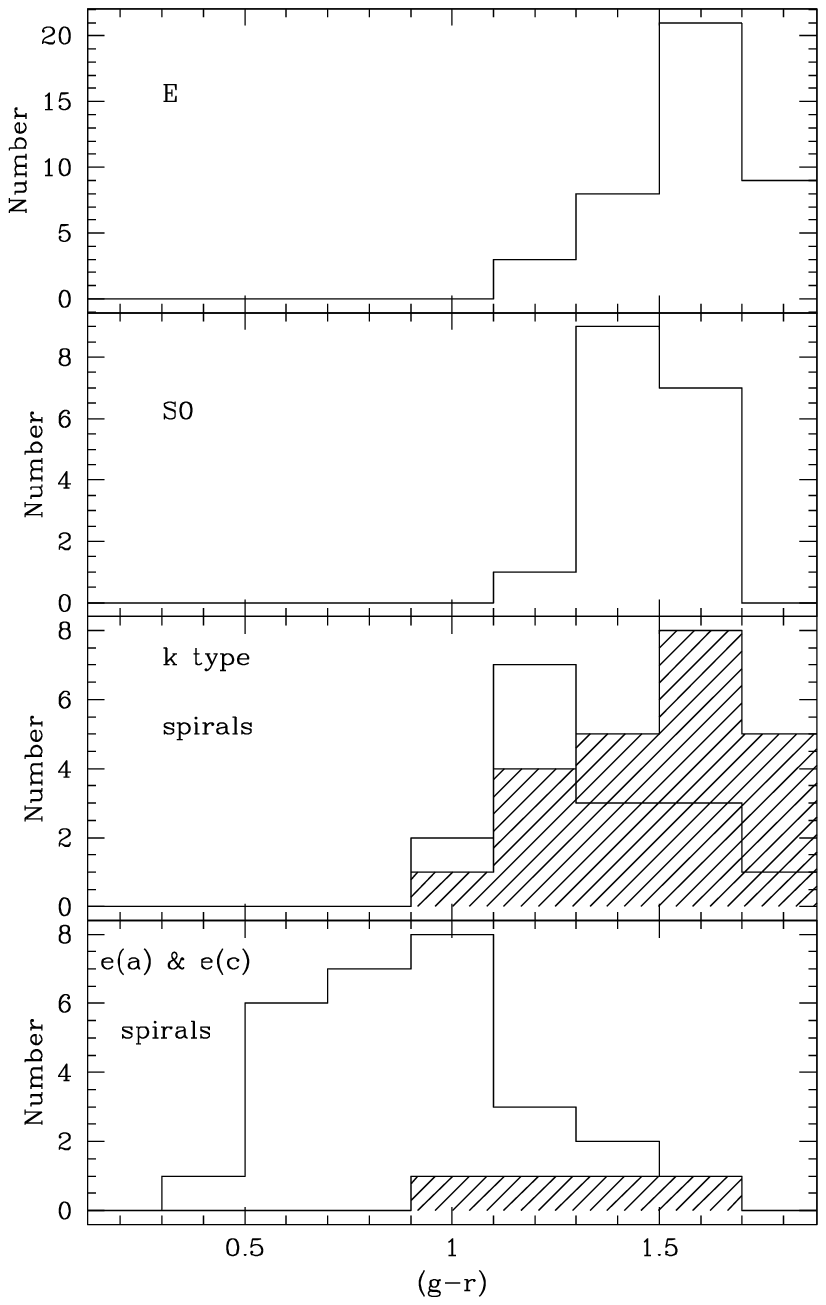,angle=0,width=4.5in}}
\vspace{-1.2in}
\noindent{\scriptsize
\addtolength{\baselineskip}{-3pt} 
\hspace*{0.3cm}
Fig.~8.\ The color distributions of the k--type spirals as compared
with other Hubble and spectral classes: k--type ellipticals (top
panel); k--type S0's (second panel from the top); k--type spirals of Sa
type (shaded area) and Sb--later types (empty histogram) (third panel);
spirals with emission (e(a) and e(c) classes) of Sa's (shaded area) and
Sb--later types (empty histogram) in the bottom panel.
 
\addtolength{\baselineskip}{3pt}
}
\smallskip

Dressler et al.\ (1997) found that while the fraction of ellipticals in
the  clusters of this sample is comparable to that in low-redshift
clusters, the fraction of S0 galaxies is 2--3 times  lower than
expected,  with a corresponding increase in the spiral fraction.   The
k+a/a+k galaxies  are the best candidates for S0 progenitors because of
both their spectrophotometric characteristics and the predominance of
disk-like morphologies.  The fact that the k+a/a+k galaxies are mostly
spirals -- i.e.\ {\it the spiral arms in their disks are still visible
} -- shows that the cessation of the star formation does not produce
{\it at the  same time} a change in morphology and that either two
timescales or two different physical processes cause the two observed
transformations, the halting of the star formation and the
production of S0's.  If this is the case, then the timescale of the
`morphological evolution' (from spirals to S0's) must be longer than
that of the spectro-photometric evolution (about 1\,Gyr).  The fact
that S0's and ellipticals have indistiguishable color--magnitude
relations in these distant clusters (Ellis et al.\ 1997) would be a
natural consequence of this delayed morphological transformation.  This
issue will be discussed in detail in a forthcoming paper (Couch et
al.\ in preparation).

\subsection{Cluster to cluster variations}

Our sample includes clusters with a wide range in richness, mass and
X-ray luminosity (Smail et al.\ 1997a). We can therefore investigate
the connection between star formation activity in the galaxy
populations and the global cluster properties.  Such connections could
shed some light on the physical mechanism(s) responsible for the strong
evolution in the galaxy populations within rich clusters  between
$z\sim 0.4$ and the present day.

We start by noting that  all the clusters, except 3C\,295,  contain
galaxies across the whole range of spectral classes (Table~4).  The
exception in 3C\,295 is that there are no galaxies with e(b) spectra
observed; however this absence is not statistically significant.
Overall the clusters show roughly comparable galaxy populations.

Although the spectral populations of the clusters show little
variation,  we wish to investigate whether the proportion of the
active  population of the clusters which are in a post-starburst phase
correlates with any property of the cluster.  The spectroscopic limit
is similar in all the clusters, approximately $M_V=-19+5 \, \log_{10}h$
(D99), but it should be kept in mind that the level of completeness
varies and this could introduce some luminosity effects.  X-ray
luminosities,  estimates of the central masses from lensing, and the
integrated luminosity of the whole cluster galaxy population are
available from Smail et al.\ (1997a), and the concentration index, $C$,
and spiral fractions for the clusters from Dressler et al.\ (1997).

We compare the spectral properties of the centrally concentrated
`regular' clusters (3C\,295, Cl\,0016+16, Cl\,0024+16) with those of
the less concentrated, irregular clusters (Cl\,0303+17, Cl\,0939+47,
Cl\,1447+23, Cl\,1601+43).   Interestingly, the spectral
characteristics do not seem to depend on the dynamical state of the
cluster.  We analyze the proportion of galaxies with emission lines
(em/all), the fraction of `active' galaxies act/all, (where the active
population is defined to include  the following classes: k+a/a+k, e(a),
e(c), e(b), e), and the proportion of  k+a/a+k galaxies as a fraction
of the total number of `active' galaxies (k+a,a+k/act); this latter
quantity is  a measure of the  efficiency of environmental processes in
stopping any star formation.  No significant variation is found between
the regular and the irregular clusters, when divided in terms of their
concentration.  The ratio k+a,a+k/act is 0.41$\pm$0.07 for the high
concentration clusters compared to  0.39$\pm$0.06 for the low
concentration clusters.  For em/all we find 0.29$\pm$0.04 for high
concentration  and 0.34$\pm$0.04 for low concentration, and for act/all
we find 0.49$\pm$0.05  versus 0.55$\pm$0.05 for high and  low
concentration clusters, respectively. This result is especially
interesting when compared  to the conclusions of Dressler et al.\
(1997), who found in the centrally concentrated clusters  a strong
morphology--local galaxy density relation  which is nearly absent in
the irregular clusters. This again seems to suggest that the star
formation histories and the morphological properties  may be separately
affected by the clusters  and in two different ways.

As regards the other global properties of the clusters (X-ray
luminosities, masses, optical luminosities, spiral fractions),  no
definitive conclusion can be reached on the basis of their correlation
with the proportion of the active population  which is in a
post-starburst phase.  A larger sample of clusters with a range of
properties (X-ray luminosities, masses, concentration indices) is
needed in order to conclusively disentangle the mechanisms at work in
producing the post--starburst population.  We believe that such
additional data, when combined with the kind of studies we have done in
our relatively small sample, could be decisive in identifying the
relevant mechanisms and processes.

\section{Conclusions}

By combining high quality spectroscopic and morphological information
on large samples of distant cluster and field galaxies we are finding
evidence for two transformations of the cluster populations.  The
quicker and more striking transformation is associated with the
wide-spread suppression of star formation in galaxies within the
cluster environment.  We have observed direct evidence for this process
in the spectra of galaxies within our clusters, including populations
of post-starburst (k+a/a+k) galaxies and passive (k) spirals. We
further suggest that the bulk of the progenitors of the post--starburst
population come from the emission line galaxies with strong Balmer
absorption (e(a)), which we interpret as dusty starbursts.  We suggest
the active populations represent recently accreted field galaxies.
Assuming that the mechanism which suppresses the star formation is
prompt, this suggests that at least 12\% of the luminosity of the
cluster core has been acquired in the last 1 Gyr. The suppression
process must act solely upon the gas content of the galaxy without
strongly affecting its morphological characteristics.  We suggest that
the process most likely to be responsible for the spectral evolution is
ram-pressure stripping of the galaxy's gas supply by the cluster ICM.
Gas exhaustion due to star formation and supernova-driven winds are
other possible mechanisms causing a dearth of ISM which are expected to
be at work in strong starbursts; since these processes are not directly
connected with the cluster environment, it is difficult to understand
why they would not produce a large k+a/a+k population also in the
field, unless they act in combination with a cluster-related
mechanism.

The second transformation -- which appears to occur on a longer
timescale -- is harder to directly observe, but is responsible for the
formation of the dominant S0 population in local clusters, a group
which is deficient in the  distant clusters we have studied (Dressler
et al.\ 1998).  We suggest that this process is involved in the
morphological conversion of the large populations of  post--starburst
and passive mid-type spiral galaxies  we see in these clusters.
Whether this morphological transformation is a consequence of the same
mechanism causing the quenching of star formation or whether another
physical process is responsible for it (such as `galaxy harassment',
Moore et al.\ 1996, 1998), remains a fundamental question.  Indeed it is
not obvious at this time whether any process other than passive fading
of the disk light is needed to explain the morphological evolution of
the disk galaxies.

Why should the infalling population be so evident at $z=0.4$ and not at
$z=0$?  The evolution in the field population surely plays a role:
spectroscopic surveys of field galaxies have found that the star
formation density is higher at $z=0.4$ (Lilly et al.\ 1996, Ellis at
al.\ 1996), i.e.\ more galaxies form stars more vigorously. The infall
rate is also expected to increase at higher redshifts (Bower 1991 and
private communication; Kauffmann 1995a, 1995b; Tormen 1998).  Both of
these effects will enhance the numbers of strongly star-forming
galaxies encountering the cluster environment. When combined with
possible evolution in the properties of the cluster ICM this may result
in substantially more activity in the distant clusters.

We now summarise the main conclusions of this work:

\noindent{$\bullet$} The class of cluster members with a very strong 
[O{\sc ii}] emission line (e(b), about 5\% of the cluster sample)
is mainly composed of low--luminosity, low--metallicity, late--type
galaxies. They do not evolve into any other spectral class visible in
our sample, and if their star formation is halted at some point,
by z=0 they will display the spectrophotometric properties 
of passive dwarfs.

\noindent{$\bullet$} Spectra with strong Balmer lines in absorption and
[O{\sc ii}] in emission (about 10\% of our cluster sample) cannot be
reproduced by models with a regular, continuous star formation rate.
A comparison with similar low--$z$ spectra suggests that these are dusty
galaxies whose star formation has been underestimated as a consequence
of the extinction.  These therefore are likely to be starburst galaxies
and thus represent the best candidates for progenitors of the numerous
post--starburst galaxies present in distant clusters.   The low--$z$
examples are associated with merging and strongly interacting galaxies;
from the {\it HST} images of the distant clusters, a merger or strong
interaction can clearly be associated with an e(a) spectrum in about
half of the cases.

\noindent{$\bullet$} The main difference between the cluster and the
field populations at $z \sim 0.4$--0.5 lies in the presence of a large
number of post--starburst galaxies in the clusters (see the Conclusions
of D99).  In contrast the population of dusty starbursts is numerous in
both environments, although we have found possible weak evidence for an
enhancement in the clusters. The differences between cluster and field
are consistent with a scenario in which the `active' galaxies
(post--starburst and all emission line galaxies) have been recently
acquired by the clusters, whose most evident effect is the quenching of
star formation.  In the field the fraction of spirals with spectra
similar to their low--$z$ counterparts is higher than in the cluster
(54\%), with a considerable incidence of galaxies with enhanced star
formation (24\%).

\noindent{$\bullet$} Only 10\% of the cluster members morphologically
classified as spirals have spectra similar to low--redshift typical
spirals, while the great majority have either an enhanced or a
suppressed  star formation rate.  We have identified a population of
cluster spirals with  `passive' spectra and red colors and discussed
their probable association with at least a fraction of the
post--starburst, spiral population.  The fact that most of the
post--starburst galaxies display spiral morphologies indicates that
either the timescale or the process responsible for halting the star
formation must be different from the one that causes the morphological
transformation.  The passive spirals we observe could either
subsequently evolve into passive S0's, or preserve their morphological
and spectrophotometric properties until z=0. Detailed studies are
needed to establish the incidence of passive spirals in low--$z$
clusters.

\noindent{$\bullet$} A summary of results from all the aspects of the
MORPHS program dealing with the evolution of the galaxy populations in
distant clusters will be presented in Couch et al.\ (in preparation).

\section*{Acknowledgements} 

We wish to thank C.\ Liu and R.\ Kennicutt for providing us their
spectrophotometric atlas of merging galaxies and D.\ Zaritsky and E.D.\
Skillman for sending us their data in a convenient form.  This work has
greatly benefited from discussions with R.\ Terlevich, F.\ Governato,
S.\ Ettori, C.\ Rola, D.\ Carter, R.\ Bower, J.\ van Gorkom.  We also
wish to thank Steven Allen for useful discussions and for computing the
K--corrections of the X-ray luminosities.  We acknowledge the
availability of the Kennicutt's (1992) atlas of  galaxies and the
Jacoby et al.'s stellar library from the NDSS-DCA  Astronomical Data
Center.  This work was supported in part by the  Formation and
Evolution of Galaxies network set up by the European  Commission under
contract ERB FMRX-CT96-086 of its TMR program.  This research has made
use of the  NASA/IPAC Extragalactic Database (NED) which is operated by
the Jet Propulsion Laboratory, Caltech, under contract with the
National Aeronautics and Space Administration. IRS, RSE and RMS
acknowledge support from the Particle Physics and Astronomy Research
Council.  AD and AO acknowledge support from NASA through STScI grant
3857.    WJC acknowledges support from the Australian Department of
Industry, Science and Technology, the Australian Research Council and
Sun Microsystems.

\section*{Appendix A: metallicities of e(b) galaxies}

There are various calibrations of the $R_{23}$ index,  based on direct
measurements of  the electron temperature for low metallicity regions,
and on  photoionization models for high metallicity regions.  We adopt
here the calibration given by Zaritsky et al.\ (1994), which is an
average of three different calibrations:

\begin{equation}
12 + \log_{10}(O/H) = 9.265 - 0.33 x -0.202 x^2 - 0.207 x^3 - 0.333 x^4
\end{equation}

where $x=\log_{10}(R_{23})$.  The $R_{23}$ index is especially valuable
to estimate {\it  relative} abundances, while the absolute calibration
is more uncertain, typically 0.2 dex (Zaritsky et al.\ 1994).

The solar value is $(12+ \log_{10}(O/H))=8.93$, $\log_{10}R_{23} \simeq 0.6$
according to equation (1) and it is shown as a dotted line in Fig.~9.
H{\sc ii} regions in giant field spirals at low redshift have [O/H] between
+0.1 and +0.4 solar, typically around +0.3  (Zaritsky et al.\ 1994),
corresponding to a metallicity of 1.2--2.0 $Z/Z_{\odot}$.

\hbox{~}
\vspace*{-0.5in}
\centerline{\psfig{file=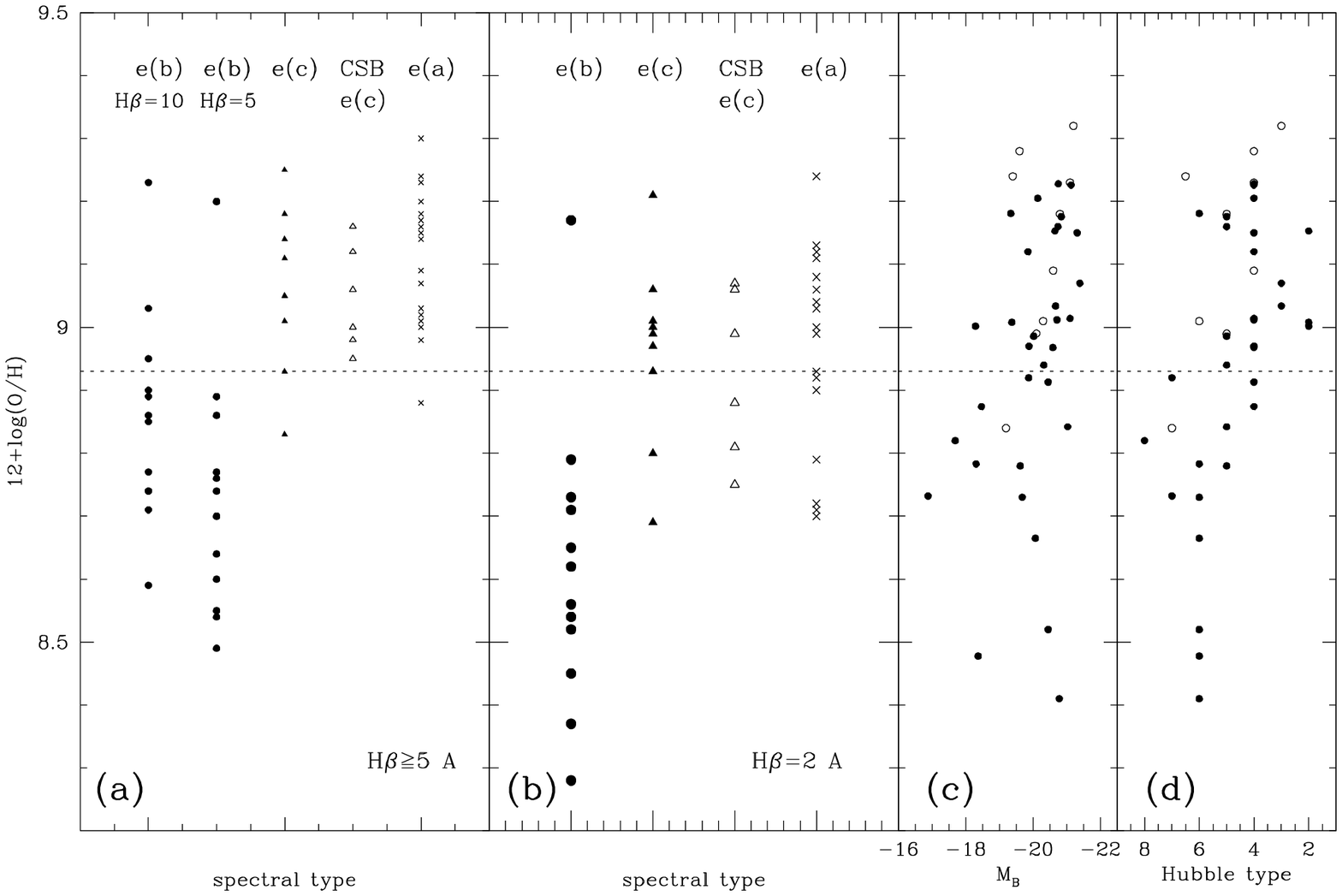,angle=0,width=3.4in}}
\vspace*{-0.3in}
\noindent{\scriptsize
\addtolength{\baselineskip}{-3pt} 
\hspace*{0.3cm} Fig.~9.\ 
Panels a) and b) show our distant cluster results as a function of the
spectral type for different $\rm H\beta$ absorption corrections:  in
panel a) ${\rm H\beta}_{abs}= 5$\AA, except in the case of e(b) (${\rm
H\beta}_{abs}=10$) and for the e(a) galaxies with $\rm H\delta > 5$\AA
(${\rm H\beta}_{abs}=\rm H\delta$); in panel b) the adopted ${\rm
H\beta}_{abs}$ is 2\AA, the same as in panels c) and d).  Panels c) and
d) show the data for low redshift field spirals from Zaritsky et
al.\ (filled dots) and Virgo spirals from Skillman et al.\ (1996)
(empty dots) as a function of the absolute $B$ magnitude and of the T
type.  The plotted values are characteristic abundances measured at  a
given fraction of the isophotal radius (0.4 $R_0$) and can be
considered as representative of an `integrated abundance' (Kobulnicky
private communication and in preparation, see also  Kobulnicky \&
Zaritsky 1998 in preparation).  The data show that the mean abundance
increases with the luminosity and for earlier Hubble types; this
luminosity--metallicity relation is known to  be valid to much fainter
magnitudes, spanning over 10 magnitudes in $M_B$ (not shown, see
Zaritsky et al.\ 1994).  Skillmann et al.\ (1996) and previous works
found that the most HI deficient Virgo spirals have {\it larger} mean
abundances than peripheral and field galaxies of comparable luminosity
or Hubble type.  Note that panels c) and d) cannot be directly compared
with panel a) because they have different ${\rm H\beta}_{abs}$.

\addtolength{\baselineskip}{3pt}
}
\smallskip

The $R_{23}$ index could be determined for 14 spectra  out of the 20
e(b)--type cluster members in D99.  For comparison purposes, we have
also measured the index in the other  spectral classes, for any
instances where the $\rm H\beta$ line was found in emission;  this was
possible in 9 e(c), 6 CSB--e(c) (the bluest examples of e(c) galaxies,
D99) and 17 e(a) spectra.  

In our $F_{\nu}$ flux-calibrated spectra (see D99), we measured the
line fluxes applying a correction factor 1.753 to the [O{\sc ii}] flux
to convert it into an $F_{\lambda}$--flux relative to the [O{\sc iii}]
and $\rm H\beta$ lines.  In the case of the uncalibrated spectra in
D99, the instrumental response as a function of $\lambda$ was found
from the comparison  between the k--type spectra from D99 and an
`elliptical' model  (exponentially decaying SFR with a short timescale
$\tau =0.5$\,Gyr) at each cluster redshift.  The correction factor for
the [O{\sc ii}] flux relative to [O{\sc iii}] was taken to be the ratio
of the factors at the  two observed wavelengths.  This choice provides
a lower limit for the [O{\sc ii}] flux and therefore an upper limit on
the metallicities.  A comparison between the index ranges of each given
spectral type  as derived from fluxed and uncalibrated spectra shows
good agreement,  confirming the validity of the correction applied to
the uncalibrated spectra.

Before computing the index it is necessary to correct the  observed
$\rm H\beta$ flux for the underlying stellar absorption.  The results
shown in panel a) of Fig.~9 have been found applying a 5\AA\ correction
or larger.  The 5\AA\ value seems the most reasonable for the e(c)
spectra, based on modeling of spiral spectra and the other  observed
Balmer lines in spirals.  In Fig.~9 the e(c) and the `CSB--e(c)'
classes (very blue galaxies with an e(c) spectrum, D99) have been kept
separated in order to study possible differences in metallicity.  The
e(a) spectra have by definition strong Balmer lines in absorption and
therefore in panel a) we have chosen  ${\rm EW(H\beta)}_{abs}=\rm
EW(H\delta)$ if $\rm EW(H\delta)>5 $\AA\ and EW(${\rm
H\beta})_{abs}=5$\AA\   otherwise.  In the case of the e(b) spectra we
computed the index both for  ${\rm H\beta}_{abs}=5$\AA\ and for
${\rm H\beta}_{abs}=10$\AA, which is an appropriate limit for
starburst galaxies (R.\ Terlevich, priv.\ comm.).  Overestimating  the
correction for absorption (i.e.\ overestimating the $\rm H\beta$ flux
in emission) results in an overestimate of the metallicities and
therefore the results in panel a) can be regarded as upper limits  on
the absolute metal content.

The metallicities in e(b) H{\sc ii} regions are clearly lower than
those  in all the other emission spectral classes in our sample,
\footnote{While the $R_{23}$ indexes (and therefore the metallicities)
measured for the e(b) spectra can be considered as typical of this
spectral class in our  sample, it is uncertain if the results for the
e(c), CSB and e(a) classes are representative of the whole sample of
that given class: our spectra do  not allow a reliable estimate of weak
$\rm H\beta$ fluxes in emission, therefore we had to restrict this
analysis to galaxies with a clear $\rm H\beta$ line in emission.  These
could represent the `most active' galaxies in each class and therefore
probably the most metal poor cases, if the usual metallicity--star
formation relations are valid at this redshift.

The sample is too small for a conclusive comparison between CSB and
non-CSB e(c) spectra, but suggests that the metallicities of the former
class are  more similar to those of the latter than to the e(b)
galaxies.  The relatively high $Z$ of CSB galaxies favors the
hypothesis that the blue colors are due to strong star formation, as
opposed to very low metallicities.} and this result is valid even for
the most conservative choice of ${\rm H\beta}_{abs}$ (10 \AA).  The
H{\sc ii} regions in e(b) galaxies generally have lower than solar
metallicity and display a wide range in $Z$ ($-0.5$ dex solar to solar,
$Z/Z_{\odot}$ between 1/3 and 1, in the case $H{\beta_{abs}=5}$\AA).

We next wish to compare the metallicities of our distant cluster
galaxies with the data of nearby field and cluster spirals (Zaritsky et
al.\ (1994)  and Skillmann et al.\ (1996), panels c) and d) in
Fig.~9).  The low--$z$ spectra have been corrected with a fixed 2\AA\
${\rm H\beta}_{abs}$ and panel b) of Fig.~9 presents the metallicities
of our sample found with the same correction.  When comparing with the
low-redshift data, one should bear in mind that the galaxies in our
sample for which the $\rm H\beta$ line could be  measured in emission
(all galaxies in panels a) and b) of Fig.~9) belong to late Hubble
types, typically Sbc or later (T types $>$ 4); the e(b) subsample is
mainly composed of Sd/Irr galaxies (T types 7--10, D99). We also notice
that the spectra in our sample are not corrected for reddening while
the low redshift spectra are, and our measures are therefore upper
limits on the metallicities.

~From Fig.~9 we conclude that the the H{\sc ii} region metallicities of
the e(b) galaxies are significantly lower than those of any other
spectral class and that these low abundances are  comparable to those
in low luminosity, very late type spirals and the most luminous
irregulars at low redshift.

\smallskip


\begin{thebibliography}{}
\itemsep=0in

\bibitem{}
Abraham, R.\,G., Smecker-Hane, T.\,A., Hutchings, J.\,B., Carlberg, R.\,G.,
Yee, H.\,K.\,C., Ellingson, E., Morris, S., Oke, J.\,B., Rigler, M., 
1996, ApJ, 471, 694

\bibitem{}
Balogh, M.\,L., Morris, S.\,L., Yee, H.\,K.\,C., Carlberg, R.\,G., Ellingson, E., 1997,
ApJ, 488, L75

\bibitem{}
Balogh, M.\,L., Schade, D., Morris, S.\,L., Yee, H.\,K.\,C., Carlberg, R.\,G.,
Ellingson, E., 1998, ApJ, submitted (astro-ph 9806146)

\bibitem{}
Barbaro G. \& Poggianti B.\,M., 1997, A\&A, 490, 504

\bibitem{}
Barger A.\,J., Aragon-Salamanca A., Ellis R.\,S., Couch W.\,J., Smail I.,
Sharples R.\,M., 1996, MNRAS, 279, 1

\bibitem{}
Barger, A.\,J., Arag\`on-Salamanca, A., Smail, I.,
Ellis, R.\,S., Couch, W.\,J., Dressler, A., Oemler, A.\ Jr, Poggianti, B.\,M, 
Sharples, R.\,M., 1998, ApJ, 501, 522 

\bibitem{}
Belloni P., Bruzual A.\,G., Thimm G.\,J., Roser H.-J., 1995, A\&A, 297, 61

\bibitem{}
Bower, R.\,G., 1991, MNRAS, 248, 332

\bibitem{}
Bower, R.\,G., Lucey, J.\,R., Ellis, R.\,S., 1992, MNRAS, 254, 601

\bibitem{}
Butcher, H., Oemler, A.\ Jr. 1978, ApJ, 226, 559

\bibitem{}
Butcher, H., Oemler, A.\ Jr. 1984, ApJ, 285, 426

\bibitem{}
Byrd, G., Valtonen, M., 1990, ApJ, 350, 89

\bibitem{}
Carter D., Prieur J.\,L., Wilkinson A., Sparks W.\,B., Malin D.\,F., 1988, MNRAS,
235, 813

\bibitem{}
Couch W.\,J., Barger A.\,J., Smail, I., Ellis R.\,S., Sharples R.\,M., 1998,
ApJ, 497, 188

\bibitem{}
Couch, W.\,J., Ellis, R.\,S., Sharples, R.\,M., Smail, I., 1994, ApJ, 430,
 121

\bibitem{}
Couch W.\,J., Sharples R.\,M., 1987, MNRAS, 229, 423 (CS87)

\bibitem{}
Dressler A., Gunn J.\,E., 1982, ApJ, 263, 533

\bibitem{}
Dressler A., Gunn J.\,E., 1983, ApJ, 270, 7

\bibitem{}
Dressler A., Gunn J.\,E., 1992, ApJS, 78, 1 

\bibitem{}
Dressler, A., Oemler, A.\ Jr., Butcher, H., Gunn, J.\,E., 1994, ApJ, 430, 107

\bibitem{}
Dressler, A., Oemler, A.\ Jr., Couch, W.\,J., Smail, I.,
Ellis, R.\,S., Barger, A., Butcher, H., Poggianti, B.\,M., Sharples,
R.\,M., 1997, ApJ, 490, 577

\bibitem{}
Dressler, A., Smail, I., Poggianti, B.\,M., Butcher, H., Couch, W.\,J.,
Ellis, R.\,S., Oemler, A.\ Jr., 1999, ApJS, in press (D99)

\bibitem{}
Duc, P.\,A., Mirabel, I.\,F., Maza, J., 1997, A\&AS, 124, 533

\bibitem{}
Edmunds, M.\,G. \& Pagel, B.\,E.\,J., 1984, MNRAS, 211, 507

\bibitem{}
Ellis, R.\,S., Colless, M., Broadhurst, T.\,J., Heyl, J.\,S., Glazebrook, K.,
1996, MNRAS, 280, 235

\bibitem{}
Ellis, R.\,S., Smail, I., Dressler, A., Couch. W.\,J., Oemler, A.\ Jr.,
Butcher, H., Sharples, R.\,M., 1997, ApJ, 483, 582

\bibitem{}
Fabricant, D.\,G., McClintock, J.\,E., Bautz, M.\,W.,  1991, ApJ, 381, 33

\bibitem{}
Fabricant, D.\,G., Bautz, M.\,W., McClintock, J.\,E.,  1994, AnJ, 107, 8

\bibitem{}
Fisher, D., Fabricant, D., Franx, M., van Dokkum, P., 1998, ApJ, 498, 195

\bibitem{}
Gallego, J., Zamorano, J., Rego, M., Vitores, A.\,G., 1997, ApJ, 475, 502

\bibitem{}
Gunn, J.\,E., Gott, J.\,R., 1972, ApJ, 176, 1

\bibitem{}
Hammer, F., Flores, H., Lilly, S.\,J., Crampton, D., Le Fevre, O., Rola, C.,
Mallen-Ornelas, G., Schade, D., Tresse, L., 1997, ApJ, 481, 49

\bibitem{}
Hammer, F., Flores, H., 1998, astro-ph 9806184

\bibitem{}
Henry, J.\,P., Lavery, R.\,J., 1987, ApJ, 323, 473

\bibitem{}
Jacoby, G.\,H., Hunter, D.\,A., Christian, C.\,A., 1984, ApJS, 56, 257

\bibitem{}
Kauffmann, G., 1995a, MNRAS 274, 153

\bibitem{}
Kauffmann, G., 1995b, MNRAS 274, 161

\bibitem{}
Kenney, J.\,D.\,P., Young, J.\,S., 1986, ApJL 301, L13

\bibitem{}
Kenney, J.\,D.\,P., Young, J.\,S., 1989, ApJ 344, 171

\bibitem{}
Kennicutt, R.\,C.\ Jr, 1992a, ApJS, 79, 255

\bibitem{}
Kennicutt, R.\,C.\ Jr, 1992b, ApJ, 388, 310

\bibitem{}
Kennicutt, R.\,C.\ Jr, 1998, ARAA, 36 (astro-ph 9807187)

\bibitem{}
Kim, D.\,C., Veilleux, S., Sanders, D.\,B., 1998, ApJ, in press (astro-ph 9806149)

\bibitem{}
Koo, D.C., Guzm\'an R., Gallego J., Wirth, G.D., 1997, ApJ, 478, L49

\bibitem{}
Koopmann, R.\,A., Kenney, J.\,D.\,P., 1998, ApJ, 497, L75

\bibitem{}
Kurucz, R., 1993, CD-ROM n.\ 13, version 22/10/93

\bibitem{}
Lavery, R.\,J., Henry, J.\,P., 1986, ApJ, 304, L5

\bibitem{}
Lavery, R.\,J., Henry, J.\,P., 1988, ApJ, 330, 596

\bibitem{}
Lilly, S.\,J., LeFevre, O., Hammer, F., Crampton, D., 1996, ApJL, 460, L1

\bibitem{}
Liu C.T., Kennicutt R.\,C. Jr, 1995a, ApJS, 100, 325

\bibitem{}
Liu C.T., Kennicutt R.\,C. Jr, 1995b, ApJ, 450, 547

\bibitem{}
Moore, B., Katz, N., Lake, G., Dressler, A.,  Oemler, A.\ Jr. 1996,
Nature, 379, 613

\bibitem{}
Moore, B., Lake, G., Katz, N., 1998, ApJ, 495, 139

\bibitem{}
Morris S.\,L., Hutchings J.\,B., Carlberg R.\,G., Yee H.\,K.\,C., Ellingson E.,
Balogh M.\,L., Abraham R.\,G., Smecker-Hane T.\,A., 1998 (astro-ph 9805216)

\bibitem{}
Newberry M.\,V., Boroson T.\,A., Kirshner R.\,P., 1990, ApJ, 350, 585

\bibitem{}
Oemler, A.\ Jr., Dressler, A.,  Butcher, H., 1997, ApJ, 474, 561

\bibitem{}
Osterbrock, D.\,E., 1989, {\it  Astrophysics of Gaseous Nebulae and Active 
Galactic Nuclei, } (Mill Valley, California:University Science Books).

\bibitem{}
Pagel, B.\,E.\,J., Edmunds, M.\,G., Blackwell, D.\,E., Chun, M.\,S., Smith, G.,
1979, MNRAS, 189, 569

\bibitem{}
Poggianti B.\,M., 1994, PhD Thesis, University of Padova

\bibitem{}
Poggianti B.\,M. \& Barbaro G., 1996, A\&A, 314, 379

\bibitem{}
Poggianti B.\,M. \& Barbaro G., 1997, A\&A, 325, 1025

\bibitem{}
Sandage, A., 1986, A\&A, 161, 89

\bibitem{}
Sanders, D.\,B., Mirabel, I.\,F., 1996, Ann.\ Rev.\ Astr.\ Astroph., 34, 749

\bibitem{}
Schommer, R.\,A., Bothun, G.\,D., 1983, ApJ, 88, 577

\bibitem{}
Skillman, E.\,D., Kennicutt, R.\,C. Jr., Shields, G.\,A., Zaritsky, D., 1996, ApJ,  462, 147

\bibitem{}
Smail, I., Ellis, R.\,S., Dressler, A., Couch, W.\,J., Oemler,
A.\ Jr, Butcher, H., Sharples, R.\,M., 1997a, ApJ, 479, 70 

\bibitem{}
Smail, I., Dressler, A., Couch, W.\,J., Ellis, R.\,S., Oemler,
A.\ Jr, Butcher, H., Sharples, R.\,M., 1997b, ApJS, 110, 213 (S97)

\bibitem{}
Smail, I., Ivison, R.J., Blain, A.W., Kneib, J.-P.,
1998, ApJL, submitted

\bibitem{}
Soucail, G., Mellier, Y., Fort, B., Cailloux, M., 1988, A\&AS, 73, 471

\bibitem{}
Stasinska, G., 1990, A\&AS, 83, 501

\bibitem{}
Tormen, G., 1998, MNRAS, 297, 648

\bibitem{}
van den Bergh, S., 1976, ApJ, 206, 883

\bibitem{}
van den Bergh, S., 1991, PASP, 103, 390

\bibitem{}
van Dokkum, P.\,G., Franx, M., Kelson, D.\,D., Illingworth, G.\,D., Fisher,
D., Fabricant, D., 1998, ApJ, 500, 714

\bibitem{}
Wilson, G., Smail, I., Ellis,
R.\,S., Couch, W.\,J., 1997, MNRAS, 284, 915

\bibitem{}
Wirth, G.\,D., Koo, D.\,C., Kron, R.\,G., 1994, ApJ, 435, L105

\bibitem{}
Wu, H., Zou, Z.\,L., Xia, X.\,Y., Deng, Z.\,G., 1998, A\&A, in press (astro-ph
9804068)

\bibitem{}
Zabludoff A.\,I., Zaritsky D., Lin H., Tucker D., Hashimoto Y., Shectman S.A.,
Oemler A., Kirshner R.\,P., 1996, ApJ, 466, 104

\bibitem{}
Zaritsky D., Kennicutt R.\,C. Jr., Huchra J.\,P., 1994, ApJ, 420, 87

\end{thebibliography}
\end{document}